\documentclass[twocolumn,showpacs,preprintnumbers,amsmath,amssymbm,prd]{revtex4}

\usepackage{graphicx}
\usepackage{dcolumn}
\usepackage{bm}
\usepackage{epsf}
\usepackage{amsfonts}

\newcommand{\beq}{\begin{equation}}
\newcommand{\eeq}{\end{equation}}
\newcommand{\beqr}{\begin{eqnarray} \nonumber}
\newcommand{\eeqr}{\end{eqnarray}}
\newcommand{\beqrb}{\begin{eqnarray}}
\newcommand{\eeqrb}{\nonumber \end{eqnarray}}

\newcommand{\fin}{\mbox{ .}}
\newcommand{\coma}{\mbox{ ,}}

\newcommand{\lrgspc}{\,\,\,\,\,\,\,\,\,}

\newcommand{\abs}[1]{{\lvert #1 \rvert}}
\newcommand{\R}{{\mathbb R}}

\newcommand{\Z}{{\mathbb Z}}

\newcommand{\tT}{\widetilde{\mathcal{T}}}
\newcommand{\tR}{\widetilde{\mathcal{R}}}
\newcommand{\nT}{\mathcal{T}}
\newcommand{\nR}{\mathcal{R}}

\newcommand{\re}{\mathop{\mathrm{Re}}}
\newcommand{\im}{\mathop{\mathrm{Im}}}

\newcommand{\TTM}{{\mbox{\tiny TTM}}}
\newcommand{\QNM}{{\mbox{\tiny QNM}}}
\newcommand{\TRM}{{\mbox{\tiny TRM}}}

\begin{document}

\bibliographystyle{apsrev}


\title{Asymptotic Spectroscopy of Rotating Black Holes}

\author{Uri Keshet}
\altaffiliation{Friends of the Institute for Advanced Study member}
\email{keshet@sns.ias.edu}

\author{Andrew Neitzke}
\email{neitzke@ias.edu}

\affiliation{Institute for Advanced Study, Einstein Drive, Princeton, NJ
08540, USA}

\date{\today}

\begin{abstract}

We calculate analytically the transmission and reflection amplitudes
for waves incident on a rotating black hole in $d=4$, analytically continued
to asymptotically large, nearly imaginary frequency.  These amplitudes determine
the asymptotic resonant frequencies of the black hole, including quasinormal modes,
total-transmission modes and total-reflection modes.  We identify these modes
with semiclassical bound states of a one-dimensional Schr\"{o}dinger equation, localized
along contours in the complexified $r$-plane which connect turning points of
corresponding null geodesics.
Each family of modes
has a characteristic temperature and chemical potential.  The relations between
them provide hints about the microscopic description of the black hole in
this asymptotic regime.

\end{abstract}


\maketitle

\section{Introduction}

The experimental inaccessibility of the Planck scale motivates searches for
indirect windows on the theory of quantum gravity.
Quantization of black holes could play an important role in this
regard, analogous to that of atomic models in the development
of quantum mechanics. In the search for a quantum theory of gravity, the
formation and evaporation of black holes as measured by an observer
very far from the horizon are generally assumed to be consistent with the
basic principles of general relativity and quantum mechanics.
Related classical processes, such as waves
scattering off the black hole, thus play an important role in constraining
quantum gravity.

The problem of determining the transmission (${\nT}$) and reflection (${\nR}$)
amplitudes of linearized perturbations incident from spatial infinity is
central in the study of black holes \cite{Chandrasekhar:1985kt}. Information about
the classical black hole encoded in ${\nT}$ and ${\nR}$ has been associated in some cases
with its quantum counterpart \cite{Aharony:1999ti}. However, in spite of an
intensive study of black hole spectroscopy, analytic results for $\nT$ and $\nR$ in the
general case of a rotating black hole have so far been
available only in the low-frequency limit.

Isolated, classical black holes, like most systems with radiative boundary
conditions, are characterized by a discrete set of complex ringing
frequencies $\omega(n) = \omega_R+i\omega_I$ known as quasinormal modes
(QNMs) \cite{Nollert:1999ji}. These resonances play an important
role in modeling the time evolution of black hole perturbations;
simulations show that at intermediate times they make the dominant
contribution.
The discrete QNM spectrum, given by the poles of ${\nT}$ and ${\nR}$, extends (for
fixed quantum numbers) along the imaginary $\omega$-axis to
infinitely large $|\omega_I|$, so one might suspect that the amplitudes
${\nT}$ and ${\nR}$ have an interesting structure at large, nearly imaginary frequencies.
Numerical studies have revealed a complicated, rich spectrum at low
frequencies even for a spherical black hole. Highly-damped resonances with
$|\omega_R| \ll |\omega_I|$ are known to be less sensitive to the details
of the perturbation, suggesting that ${\nT}$ and ${\nR}$ may admit a simple
interpretation in this regime. For example, it has been argued that one can read off
the quantum of area of the black hole horizon from the highly-damped
QNM frequencies \cite{Hod:1998vk}.

The transmission-reflection problem has previously been solved
analytically in the highly-damped regime for spherical black holes
\cite{Neitzke:2003mz,Harmark:2007jy}. Recently, the highly-damped QNM spectrum of rotating
black holes was analytically derived \cite{Keshet:2007nv}. Here we combine
the tools developed in \cite{Neitzke:2003mz} and in \cite{Keshet:2007nv} to
solve the highly-damped transmission-reflection problem for a
rotating black hole in four dimensions.

The resulting analytic expressions for
${\nT}$ and ${\nR}$ capture, in addition to the QNM frequencies, various other
resonances of the system.
We show that these resonances can be identified directly with
semiclassical bound states of an effective one-dimensional wave equation.
They live naturally along steepest-descent (anti-Stokes) contours between two complex turning
points of corresponding null geodesics,
and their frequencies satisfy a complex Bohr-Sommerfeld equation.
The highly-damped quasinormal modes (QNM), total-transmission modes (TTM), and
total-reflection modes (TRM) correspond to three different contours which we
interpret as ``external,'' ``internal,'' and ``mixed,'' respectively.
The resonant frequencies are $\omega(n)=\widetilde{\omega}+4\pi iT(n+\mu/4)$, where
$\Delta t=(4iT)^{-1}$ and $\widetilde{\omega}\Delta t$ are respectively the time
and angular distance elapsed along corresponding null geodesics in
the complexified black hole background, and $\mu$ is a Maslov index.

Following the philosophy of \cite{Hod:1998vk} one might hope that all of these highly-damped
resonances carry some information about the quantum theory.  One way this could happen was
proposed in \cite{Neitzke:2003mz}:  determining ${\nT}$ allows one to calculate the analytically
continued spectrum of Hawking radiation escaping from the black hole, and one can look at the result for clues
about a microscopic or ``dual'' description of the same physics, a strategy which has been
successful in other spacetimes and frequency regimes in the past \cite{Maldacena:1997ix, Maldacena:1997ih, Aharony:1999ti}.
Indeed, as we will see, our results for a rotating
black hole bear an encouraging resemblance to some examples where a dual description has been established.
There are simple relations between the parameters $T$ and
$\widetilde{\omega}$ of the three resonant modes:
\begin{align}
\frac{1}{2 \, T_{\TTM}} - \frac{1}{2 \, T_\QNM} &= \frac{1}{2 \, T_{\TRM}} = \frac{1}{T_H}, \\
\frac{\widetilde{\omega}_{\TTM}}{2 \, T_{\TTM}} - \frac{\widetilde{\omega}_{\QNM}}{2 \, T_{\QNM}} &= \frac{\widetilde{\omega}_{\TRM}}{2 \, T_{\TRM}} = \frac{m \Omega}{T_H} + 2 \pi i s.
\end{align}
Here $T_H$ is the Hawking temperature of the black hole, $\Omega$ the angular velocity of the event horizon,
$m$ the azimuthal quantum number of the perturbation, and $s$ the spin of the perturbing field.
The analytically
continued decay spectrum has a Boltzmann-like form, inversely proportional
to $e^{(\omega-\widetilde{\omega}_\QNM)/2 T_{\QNM}}+1$.
These results support the
point of view that the QNMs and TTMs correspond to distinct microscopic degrees of freedom,
which interact to produce Hawking radiation.

The paper is organized as follows.
In \S\ref{sec:Analysis} we formulate the transmission-reflection problem
for a rotating black hole, derive the amplitudes ${\nT}$ and ${\nR}$, and
determine some of the resonances. In
\S\ref{sec:Excitations} we identify the highly-damped regime as a
``classical'' limit in which the scattering problem reduces to tunneling
between neighboring contours in the complex $r$-plane, and study excitations
corresponding to each contour. \S\ref{sec:Geodesics} reinterprets the
results of \S\ref{sec:Analysis} and
\S\ref{sec:Excitations} in terms of null geodesics in the complexified black hole
spacetime.  In \S\ref{sec:DecayRate} we study the analytically-continued
decay spectrum of the black hole in search for hints of an
underlying microscopic theory, and discuss analogies with cases
previously studied.
In \S\ref{sec:Discussion} we summarize the analysis and discuss its conclusions.
Some generalizations to other black holes are presented in
Appendix \S\ref{subsec:GeneralizedTR}.

We use Planck units in which $G=c=k_B=k_C=\hbar=1$, where $k_B$ is the
Boltzmann constant and $k_C=(4\pi\epsilon_0)^{-1}$ is the Coulomb force
constant.

\section{Transmission-reflection problem}
\label{sec:Analysis}

In this section we analytically solve the problem of transmission and
reflection for a rotating black hole in the highly damped regime. The
general structure of the problem is formulated in \S\ref{subsec:TRProblem}.
After this we specialize to the case of the rotating black hole.
Some physical and mathematical background is laid out in
\S\ref{subsec:Teukolsky}-\S\ref{subsec:Stokes}; in particular,
highly-damped perturbations are shown in \S\ref{subsec:EquatorialFocusing}
to be equatorially confined. The boundary conditions are described in detail in \S\ref{subsec:Boundary}.
The results are finally derived in \S\ref{subsec:CalculatingTR}, summarized in
\S\ref{subsec:Results} and interpreted in terms of Boltzmann factors in
\S\ref{subsec:BoltzmannFactors}, where some resonances are also discussed.

\subsection{Transmission and reflection}
\label{subsec:TRProblem}

Linearized perturbations propagating in black hole
spacetimes often satisfy radial equations of the form
\begin{equation}
\label{eq:WaveEquation} \left[-\frac{\partial^2}{\partial
z^2}+V_z(z)-\omega^2\right]f(z) = 0 \coma
\end{equation}
where $z = z(r)$ is a ``tortoise'' coordinate defined such that
\begin{align} \label{eq:zGeneral}
z \sim r & \text{ as } r \to \infty \, ; \nonumber \\
z \to -\infty & \text{ as } r \to r_+ \coma
\end{align}
with $r_+$ the (outer) event horizon radius.
We require that $\im(z)/\re(z)\to 0$ as $r\to r_+$ or $r\to \infty$.

We impose the purely outgoing boundary condition at the horizon
(with respect to the physical line $r>r_+$, \textit{i.e.} signals travel only into the black hole),
\begin{equation}
\label{eq:boundary_conditions}
f \sim
\begin{cases}
e^{-i\omega z}+{\nR}(\omega)e^{i \omega z} & \text{ as } r\rightarrow
\infty \coma\,  z\rightarrow \infty\,; \\
{\nT}(\omega)e^{-i \omega z} & \text{ as } r\rightarrow r_+\coma \,
z\rightarrow -\infty\coma
\end{cases}
\end{equation}
where ${\nT}$ and ${\nR}$ are respectively the transmission and reflection amplitudes
for a wave incident from infinity.  The precise definition of these boundary conditions
is delicate, especially for complex $\omega$, and will be discussed in Section \ref{subsec:Boundary}.

Constancy of the Wronskian of the two
independent solutions of Eq.~(\ref{eq:WaveEquation}) implies a ``conservation of flux'' relation,
valid for arbitrary complex $\omega$,
\begin{equation}
\label{eq:ContinuedFluxConservation}
{\nT}(\omega)\tT(-\omega)+{\nR}(\omega)\tR(-\omega)=1\coma
\end{equation}
where $\tT$ and $\tR$ are the transmission and reflection amplitudes that
correspond to a different problem, where the $\omega$-dependent terms in
$V_z$ have been modified by $\omega \to -\omega$.
A far field analysis for real $\omega$ shows that ${\nR}(\omega)\tR(-\omega)$
is the fraction of energy reflected, so
${\nT}(\omega)\tT(-\omega)$ is the absorption (transmission) probability
\cite[see Ref.][and \S\ref{subsec:Teukolsky}]{Chandrasekhar:1985kt}.

\subsection{Teukolsky's radial equation}
\label{subsec:Teukolsky}

Consider an uncharged rotating black hole of mass $M$ and angular momentum
$J$.  Linearized, massless perturbations of the black hole are described
by Teukolsky's equation \cite{Teukolsky:1972my}. For scalar perturbations, this
equation has been generalized to accommodate a non-zero black hole
electric charge $Q$ \cite{Dudley:1978vd}; in the equations to follow,
one must take $Q = 0$ except for scalar perturbations.
The perturbation is decomposed as
\begin{equation} \label{eq:Separation}
{_s\psi_{lm}}(t,r,\theta,\phi)=e^{i(m\phi- \omega t)} {_sS_{lm}}(\theta){_sR_{lm}}(r) \coma
\end{equation}
where $(t,r,\theta,\phi)$ are Boyer-Lindquist
coordinates, and $l,m$ are angular, azimuthal harmonic indices with
$-l\leq m\leq l$.
The parameter $s$ gives the spin of the field, specializing the
analysis to gravitational ($s=-2$), electromagnetic ($s=-1$), scalar
($s=0$), or two-component neutrino ($s=-1/2$) fields.
We shall henceforth omit the indices $s,l,m$ for brevity.

With the decomposition \eqref{eq:Separation}, $R$ and $S$ obey radial and angular equations,
both of confluent Heun type
\cite{Ronveaux:Heun}, coupled by a separation constant $A$. The radial equation is \cite{Teukolsky:1972my}
\begin{align} \label{eq:TeukOriginal}
& \Delta^{-s} \frac{d}{dr} \left( \Delta^{s+1} \frac{dR}{dr} \right)
+ \bigg[ \frac{K^2-2is(r-M)K}{\Delta} \\
& \lrgspc\lrgspc\lrgspc -a^2\omega^2 +2am\omega - A + 4is\omega r
\bigg] R=0 \nonumber \coma
\end{align}
where $a \equiv J/M$ and $K\equiv (r^2+a^2)\omega-am$.
$\Delta\equiv r^2-2Mr+a^2+Q^2$ vanishes at
$r_\pm \equiv M\pm(M^2-a^2-Q^2)^{1/2}$, the outer (positive sign)
and inner (negative sign) horizons.

We now focus on the highly-damped regime, roughly the limit where $|\omega_I|$ is larger than
any other scale in the problem including $\omega_R$, $M^{-1}$ and $M/J$, holding $l$ and $m$ fixed.
In this limit we may write \cite{Berti:2004um,Berti:2005gp}
\begin{equation} \label{eq:Alm}
A(\omega_I\rightarrow -\infty) = i A_1 a \omega + O(|a\omega|^{0}) \coma
\end{equation}
with $A_1\in \R$.

Eq.~\eqref{eq:TeukOriginal} may be rewritten using Eq.~(\ref{eq:Alm}) as
\begin{equation}
\left[ \frac{\partial^2}{\partial r^2} + \omega^2 V(r)^2 \right]
\left[ \Delta^{(s+1)/2} R \right] = 0 \coma
\end{equation}
where
\begin{equation}
\label{eq:V_definition} V(r) =
\frac{\sqrt{q_0+\omega^{-1}q_1+O(|\omega|^{-2})}}{\Delta}\coma
\end{equation}
with
\begin{equation}
\label{eq:q0_definition} q_0(r) \equiv (r^2+a^2)^2-a^2\Delta
\end{equation}
and
\begin{eqnarray}
\label{eq:q1_definition} q_1(r) & \equiv & -2am(2Mr-Q^2) - i a A_1\Delta  \nonumber \\
& & +2is[r(\Delta+Q^2)-M(r^2-a^2)]  \fin
\end{eqnarray}
The $q_i$ are related to
the Kerr-Newman metric \cite[\textit{e.g.}
Ref.][]{nov-frol} by $q_0=g_{\phi\phi}\Sigma$,
$\re(q_1)=2mg_{t\phi}\Sigma$, where
$\Sigma\equiv r^2+a^2\cos^2\theta$ vanishes at the ring singularity.
Near the horizons, $q_0=(A_\pm/4\pi)^2$ and $\re(q_1)=-2am(A_\pm/4\pi)$,
where $A_\pm=4\pi(r_\pm^2+a^2)=4\pi(2Mr_\pm-Q^2)$ is the area of the outer/inner horizon.

Teukolsky's radial equation may finally be written \cite{Keshet:2007nv} in the form of
Eq.~(\ref{eq:WaveEquation}), upon
defining $f\equiv\Delta^{(s+1)/2} V^{1/2} R$ and
a (nonconventional; \textit{cf.} \cite{Chandrasekhar:1985kt}) tortoise coordinate
\begin{equation}
\label{eq:z_definition} z \equiv \int^r V(r')dr' \fin
\end{equation}
The potential $V(r)$ defined in Eq.~(\ref{eq:V_definition}) is multivalued
because of the square root.  We will choose its branch cuts such
that our analysis uses only a single Riemann sheet for $V(r)$, on which
as $r \to \infty$ we have $V(r)\to +1$ and $z\to +r$, in agreement with Eq.~(\ref{eq:zGeneral}).

Eq.~(\ref{eq:z_definition}) shows that $z(r)$ is also multivalued, with
monodromy around each of the two simple poles of $V(r)$;
this monodromy will play an important role
below.

The potential appearing in Eq.~(\ref{eq:WaveEquation}) is given by
\begin{equation} \label{eq:V_z}
V_z(z) = \frac{V''}{2V^3} - \frac{3(V')^2}{4V^4}
\end{equation}
(derivatives with respect to $r$),
and satisfies $V_z = O(\omega^0)$.  It remains finite at $r_\pm$, but diverges at the
four turning points $r_i$ defined by $V(r_i)=0$, which are essential
to the analysis.
The $O(\abs{\omega}^{-2})$ term in Eq.~\eqref{eq:V_definition} should be chosen so
that $V_z$ vanishes exponentially
as $z \to -\infty$, and
$V_z = O(z^{-2})$ as $z \to \infty$;
a straightforward choice is
\begin{equation}
\omega^{-2}\left[ a^2 m^2 +i a m s(r_+-r_-) -s^2(r_+-r_-)^2/4 \right] \fin
\end{equation}

Finally we briefly discuss the relation between the wave equation (\ref{eq:WaveEquation}) and the physical absorption
probability.  In \cite{Chandrasekhar:1985kt} it is argued that for electromagnetic and gravitational
perturbations the fraction of energy reflected is $\nR(\omega) \tR(-\omega)$.
The same result is shown for scalar perturbations in \textit{e.g.} \cite{dolan-thesis},
and for fermions in \cite{Chandrasekhar:1977kf}.
In those treatments the radial equation is formulated with a different definition
of $z$ and $f$ than we are using; our $\nR(\omega) \tR(-\omega)$ nevertheless agrees with theirs.
It follows that $\nT(\omega)\tT(-\omega)$ is the absorption probability in all these cases.

\subsection{Teukolsky's angular equation: equatorial confinement}
\label{subsec:EquatorialFocusing}

Teukolsky's angular equation is \cite{Teukolsky:1972my}
\begin{align} \label{eq:TeukolskyAngular}
& \frac{1}{\sin\theta} \frac{d}{d\theta} \left( \sin\theta
\frac{dS}{d\theta} \right) = \bigg[ -(a \omega \cos\theta)^2 \\
& \lrgspc \lrgspc + \frac{(m+s\cos\theta)^2}{\sin^2\theta} + 2a\omega s
\cos\theta - s - A \bigg] S \nonumber \fin
\end{align}
$S$ is required to be regular at the regular singular
points $\theta=0$ and $\theta=\pi$ (the poles).  This condition picks out a discrete
set of solutions $S = S_l$, known as spin-weighted spheroidal wave functions
(SWSWF), and corresponding eigenvalues $A$ \cite[for review see
Ref.][and references therein]{Berti:2005gp}. In the scalar case $s=0$,
the $S_l$ reduce to the more familiar spheroidal wave functions
(SWF) \cite{flammer}.

When $|a\omega|\to\infty$ for fixed $l$,
$A=O(|a\omega|)$ is given by Eq.~(\ref{eq:Alm}) both for $s=0$
\cite[prolate-type SWFs, see Ref.][and references therein]{MR2084854} and
$s\neq 0$ \cite{Berti:2005gp}.
Then the right side of
Eq.~(\ref{eq:TeukolskyAngular}) is dominated by the first term
sufficiently far from the poles and from the equator; when $m \neq 0$
the condition is $|m/a\omega|^2 \lesssim \cos^2 \theta \lesssim 1-|m/a\omega|^2$. Very near the poles,
the second term on the RHS takes over.  Both these terms are positive in the highly-damped regime,
so we get exponential decay/growth of $S$
everywhere except in the equatorial region.  The regular boundary
conditions at the poles then require that $S$ \emph{decays} rapidly away from the
equator.  This analysis agrees with the known behavior of the asymptotic
prolate SWFs, in which the magnitude decreases rapidly with increasing
$\lvert\cos\theta\rvert$ \cite{MR2084854}.  There is some numerical evidence that
this is also the case for the SWSWFs \cite[\textit{e.g.}, Ref.][Figure 5]{Berti:2004um}.

\subsection{Stokes and anti-Stokes lines}
\label{subsec:Stokes}

\begin{figure}[h]
\centerline{\epsfxsize=7.5cm \epsfbox{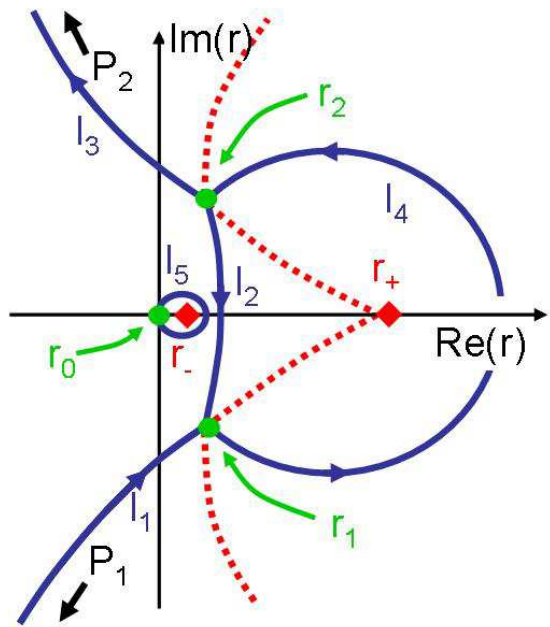}}
\caption{\label{fig:Stokes} Illustration of perturbation analysis
for a rotating black hole in the highly-damped regime. Anti-Stokes (solid)
lines, Stokes (dashed) lines and turning
points $r_i$ (disks) are shown in the complex $r$-plane for the case
$a=0.3$ and $Q=0$. Arrows along anti-Stokes lines point in the direction
of increasing $\im(z)$. Inner and outer horizon radii (diamonds) are also
shown. }
\end{figure}

Eq.~(\ref{eq:WaveEquation}) can be solved in the highly-damped regime by
evolving $f$ in the WKB approximation
\footnote{Also known as the JWKB or the Liouville-Green approximation, and
as the first order phase integral method \cite{froman-jwkb}.}
along specific contours in the complex $r$-plane.
Such a contour, consisting of anti-Stokes lines defined by $\re
(i\omega z)=0$, is constructed as follows. Let $\widetilde{r}_1$ and $\widetilde{r}_2=\widetilde{r}_1^*$ be
the two complex conjugate roots of $q_0$, with $\re (\widetilde{r}_1)>0$ and $\im (\widetilde{r}_1) < 0$.
The two other roots $\widetilde{r}_0$ and $\widetilde{r}_3$ are real;
for $Q=0$ they are $\widetilde{r}_0=0$ and $\widetilde{r}_4=-2\re(\widetilde{r}_{1,2})$.
Let $r_0$, $r_1$ and $r_2$ denote the turning
points which in the $|\omega_I| \to \infty$ limit approach $\widetilde{r}_0$, $\widetilde{r}_1$
and $\widetilde{r}_2$, respectively (see Figure \ref{fig:Stokes}). Near the turning
points, $(z-z_i)\propto (r-r_i)^{3/2}$, where $z_i\equiv z(r_i)$.
Therefore three anti-Stokes lines emanate from each turning point. Two
anti-Stokes lines connect $r_1$ to $r_2$; one (denoted $l_2$) crosses the
real axis between $r_-$ and $r_+$, while the other ($l_4$) crosses it at
$r>r_+$. The third anti-Stokes line ($l_1$) emanating from $r_1$ extends
to $P_1$, where $|P_1|\rightarrow \infty$ and $\arg (P_1)=-\pi/2$. A
similar line ($l_3$) runs from $r_2$ to $P_2$, with $|P_2|\rightarrow
\infty$ and $\arg (P_2)=+\pi/2$. A Stokes line, defined by $\im (i\omega
z)=0$, emanates between every two anti-Stokes lines of each turning point.
Figure \ref{fig:Stokes} illustrates the features relevant to the analysis in
the complex $r$-plane.

Along anti-Stokes lines, the WKB approximation
\begin{equation}
\label{eq:WKBdefinition} f(z) = \left[ c_+ f_+(z-z') + c_-
f_-(z-z')\right][1+O(|\omega|^{-1})]
\end{equation}
holds, where we defined $f_\pm(z) \equiv e^{\pm i \omega z}$,
and $z'=z(r')$ is some reference point.
We use the notation
\begin{equation} \label{eq:WKBNotation}
f(l_j)=\{c_+,c_-;r'\}
\end{equation}
to describe this solution to leading order in $|\omega|^{-1}$ along the
anti-Stokes line $l_j$. Off the anti-Stokes lines, the solution may also
be written as $c_d f_d+c_s f_s$, with $f_d,f_s \in \{f_+,f_-\}$ chosen such that $f_d$ is
exponentially large (dominant) and $f_s$ is exponentially small
(subdominant) in that region. The coefficients $c_\pm$ are approximately constant along anti-Stokes lines away from
the turning points, and mix with one another near the turning points
in a way dictated by the Stokes phenomenon \cite{froman-jwkb}.
When an anti-Stokes line is crossed, the dominant and subdominant parts exchange
roles; when a Stokes line is crossed while circling a regular turning
point at $r'$, $c_d f_d+c_s f_s$ becomes $c_d f_d+(c_s\pm i c_d)f_d$, where the
positive (negative) sign corresponds to a counterclockwise (clockwise)
rotation.

\subsection{Boundary conditions}
\label{subsec:Boundary}

Next, we implement the boundary conditions
Eq.~(\ref{eq:boundary_conditions}) for a rotating black hole. This is
slightly subtle for complex $\omega$. A rigorous way to fix the boundary
condition at $r_+$ is by specifying the monodromy of the solution there,
\textit{i.e.} requiring that $f$ is an eigenvector of the monodromy matrix
with a specific eigenvalue.
A Frobenius analysis (power series expansion) of the Teukolsky equation at
$r_+$ shows that there are two independent solutions $R(r) =
(r-r_+)^{i\omega\tau_k} \left[1 + O(r - r_+)\right]$ with $k\in\{I,O\}$
corresponding to ingoing, outgoing waves with respect to the physical
region outside the black hole (\textit{i.e.} signals travel out of, into the
black hole, see \cite{Teukolsky:1973ha}; henceforth).
These solutions have
monodromies $e^{2 \pi \omega \tau_k}$ on a clockwise rotation around
$r_+$, where
\begin{equation} \label{eq:BoundaryConditionsR}
\omega \tau_k = \frac{is}{2} \pm \left[\frac{\omega - m \Omega}{4\pi T_H} -
\frac{is}{2} \right] \coma
\end{equation}
with positive (negative) sign corresponding to ingoing (outgoing) waves
\cite{Teukolsky:1973ha}. Here,
$T_H=(r_+-r_-)/A_+$ is the Hawking temperature,
and $\Omega\equiv \Omega_+ = 4\pi a/A_+$ is the angular velocity of the (outer) event horizon.
The relation between $f$ and $R$ involves an extra factor
$\Delta^{s/2}$, which is proportional to $(r-r_+)^{s/2}$ near the horizon
and has a monodromy $e^{-\pi i
s}$ on a clockwise rotation around it. The two solutions $f_k(r)$ near
$r_+$ thus have monodromies $e^{\pm 2 \pi \omega \sigma_+}$, where
\begin{equation} \label{eq:SigmaPlus}
\omega \sigma_+ = \frac{\omega - m \Omega}{4\pi T_H} - \frac{i s}{2} \fin
\end{equation}
To leading order in $\abs{\omega}^{-1}$, this expression for the
monodromies could also have been obtained by writing the two solutions as
$e^{\pm i \omega z}$ and then using the monodromy of $z$ around $r_+$;
that would give $\sigma_+=\mbox{Res}_{r\to r_+}(V)$, as used in
\cite{Keshet:2007nv}, which indeed agrees with Eq. \eqref{eq:SigmaPlus}. The
boundary condition requiring outgoing waves at the horizon,
\begin{equation} \label{eq:BoundaryRpKerr}
f(r\rightarrow r_+) \sim {\nT}(\omega)e^{-i\omega z}, \nonumber
\end{equation}
can be defined as choosing the solution with clockwise monodromy
$\Phi_O=e^{- 2 \pi \omega \sigma_+}$.

Next we consider the boundary condition at spatial infinity.
For $\omega$ slightly off the real axis, the boundary condition at $r\to\infty$ can be
continued to a point $P$ on the complex $r$-plane, lying far from the
origin on an anti-Stokes line nearest to the real axis \citep[See
Ref.][\S2.3.7]{froman-pppi}. As $\arg(\omega)$ gradually decreases from
$0$ to $-\pi/2$, $\arg(P)$ gradually increases from $0$ to $+\pi/2$, $P$
eventually becoming nearly imaginary \cite{Motl:2003cd}. When
$\re(\omega)=m\Omega$, the anti-Stokes lines go through a discontinuous
change, signaling the presence of a branch cut at these values of $\omega$
\footnote{In this case $i\omega z\simeq i\omega\sigma_+ \ln(r-r_+)$ cannot
be imaginary near $r_+$, so anti-Stokes lines cannot spiral into $r_+$.  Note
that in \cite{Hod:2005ha} the branch cut in $\omega$ was chosen differently.}.
For $\omega_R>m\Omega$, the boundary condition can then be continued to $P_2$,
\begin{equation} \label{eq:BoundaryP2}
f(P_2) \sim e^{-i\omega z} + {\nR}(\omega)e^{+i\omega z}\coma
\end{equation}
implying that $f(l_3) = \{{\nR},1;z_2\}$ up to a multiplicative factor. For
$\omega_R<m\Omega$, the boundary condition must instead be continued to $P_1$,
so $f(l_1) = \{{\nR},1;z_1\}$ up to a multiplicative factor.

\subsection{Computation}
\label{subsec:CalculatingTR}

Below we will construct a contour which asymptotically approaches $P_1$ or
$P_2$, encloses $r_+$, and consists only of anti-Stokes lines.
When the contour reaches a turning point, it circles around it,
excluding it from the enclosed region.
The contour we use to analyze the $\omega_R<m\Omega$ case is shown in Figure
\ref{fig:Computation} below.

\begin{figure}[h]
\centerline{\epsfxsize=7.5cm \epsfbox{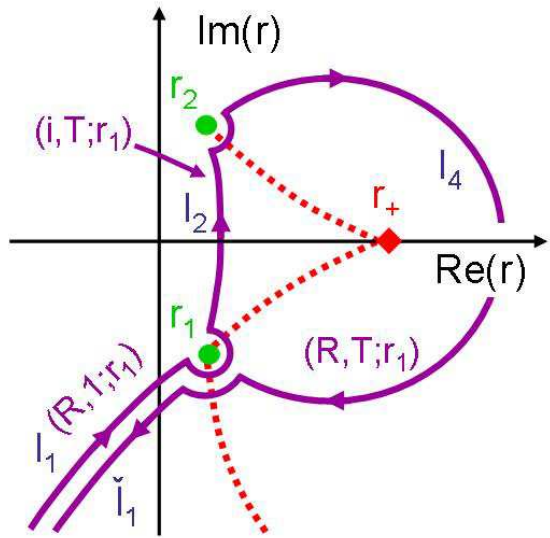}}
\caption{\label{fig:Computation}
Illustration of WKB computation for the case $\omega_R<m\Omega$. The directed
contour (solid line with arrows) encloses only the outer horizon. Bracketed
triplets denote the values of $\{c_+,c_-;r'\}$, as defined in
Eqs.~(\ref{eq:WKBdefinition})-(\ref{eq:WKBNotation}).
Other parameters and symbols are defined as in Figure \ref{fig:Stokes}. }
\end{figure}

We fix the boundary condition for a solution $f$ at $P_1$ or $P_2$ and then
evolve it along the contour in the WKB approximation.  The monodromy
along the contour must agree with that determined by
the boundary condition at $r\rightarrow r_+$; this provides a constraint
on $f$.
Furthermore, the solution strictly inside the region enclosed by $l_2$ and $l_4$ can be
approximated by $c_- f_-$ (since $f_+$ is exponentially small here).  Evaluating at
$r_+$ then gives $c_- = \nT$, while continuing to $l_{2,4}$ yields $c_- = c_-(l_{2,4})$
\cite{froman-jwkb}, so we get a second constraint, $\nT = c_-(l_{2,4})$.
These two constraints completely determine ${\nR}$ and ${\nT}$.

In the following, results accurate only to leading order in
$|\omega|^{-1}$, derived from the WKB approximation, are
indicated with the $\approx$ symbol; for fractions this generally includes corrections both
to numerator and denominator.

\subsubsection{The case $\omega_R<m\Omega$}

First consider the regime $\omega_I<0$, corresponding to time decay, and
$\omega_R<m\Omega$. Starting from $P_1$, where $f$ is given by the boundary
condition at spatial infinity,
$f(l_1) = \{{\nR},1;r_1\}$ holds along
$l_1$ till the vicinity of $r_1$. We may derive $f(l_2)$ by rotating
counterclockwise around $r_1$, from $l_1$ to $l_2$. This rotation
involves crossing two Stokes lines and the anti-Stokes line between them,
so $f(l_2)=\{i,1+i{\nR};r_1\}$. In the region enclosed by $l_2$ and
$l_4$, $f_-$ is the dominant solution; we may
therefore determine ${\nT}$ directly from $c_-(l_2)$ as
\begin{equation}
\label{eq:TvsR} {\nT}\approx 1+i{\nR} \fin
\end{equation}

Next we will follow the contour to $r_2$ along $l_2$, to $\check{r}_1$ along $l_4$,
and then to $\check{l}_1$.
Since $f$ and $z$ are multivalued functions of $r$, branched at $r_+$, traversing the
contour brings us to another Riemann sheet; we use $\check{}$ to denote objects on this
second sheet.  We first write $f(l_2)$ with $r'=r_2$, so $f(l_2) = \{i\exp
(+i\omega\delta),(1+i{\nR})\exp (-i\omega\delta);r_2\}$, where
\begin{equation} \label{eq:deltaDef}
\delta \equiv z_2-z_1 = \int_{l_2} V \, dr \fin
\end{equation}
Counterclockwise rotating around $r_2$ from $l_2$ to $l_4$ gives
\begin{eqnarray}
f(l_4)&=&
\{ie^{+i\omega\delta}+(i-{\nR})e^{-i\omega\delta},\\ \nonumber
& & (1+i{\nR})e^{-i\omega\delta};r_2\} \\ \nonumber
&=& \{[i+(i-{\nR})e^{-2i\omega\delta}] e^{2 \pi \omega \sigma_+},\\ \nonumber
& & (1+i{\nR})e^{-2 \pi \omega \sigma_+};\check{r}_1\} \fin \nonumber
\end{eqnarray}
Finally, $f(\check{l}_1)$ may be obtained by clockwise
rotating around $\check{r}_1$, thus crossing a Stokes line. This yields
\begin{align}
f(\check{l}_1)= & \{ [i+(i-{\nR})e^{-2i\omega\delta}]e^{2\pi\omega\sigma_+},
\nonumber \\
&
[1+(1+i{\nR})e^{-2i\omega\delta}]e^{2\pi\omega\sigma_+}+(1+i{\nR})e^{-2\pi\omega\sigma_+};
\nonumber \\
& \check{r}_1\} \fin
\end{align}

The only singularity of the differential equation enclosed by the contour is at $r_+$. Hence
$f(\check{l}_1)$ and $f(l_1)$ differ only by the action of the monodromy matrix at
$r_+$.  Our boundary condition requires that $f$ is an eigenvector of this monodromy
with eigenvalue $\Phi_O=\exp(-2\pi\omega\sigma_+)$.
This implies two degenerate
constraints, $c_+(\check{l}_1)/c_+(l_1) = \Phi_O$ and $c_-(\check{l}_1)/c_-(l_1)=\Phi_O$;
either one gives the same formula for ${\nR}$,
\begin{equation}
\label{eq:R_wRl0_wIl0} {\nR}(\omega) \approx i
\frac{e^{-2i\omega\delta}+1}{e^{-2i\omega\delta}+e^{-4\pi\omega\sigma_+}}
\fin
\end{equation}
Combining this result with Eq.~(\ref{eq:TvsR}) yields
\begin{equation}
\label{eq:T_wRl0_wIl0} {\nT}(\omega) \approx
\frac{e^{-4\pi\omega\sigma_+}-1}{e^{-2i\omega\delta}+e^{-4\pi\omega\sigma_+}}
\fin
\end{equation}

The same contour can be used to calculate ${\nT}$ and ${\nR}$ at
frequency $-\omega$. The only change in the analysis is due to the
reversed dominance pattern among $f_+$ and $f_-$, each becoming dominant
where it was previously subdominant and vice versa. The result
is
\begin{equation}
{\nT}(-\omega) \approx 1 \lrgspc \mbox{and} \lrgspc {\nR}(-\omega) \approx -i\fin
\end{equation}

\subsubsection{The case $\omega_R>m\Omega$}

Next, consider the regime $\omega_I<0$, $\omega_R>m\Omega$. The analysis can be
carried out exactly as in the $\omega_R<m\Omega$ case but with the contour
reflected about the real $r$-axis, and with the boundary condition at spatial
infinity continued to $P_2$.
This yields
\begin{equation}\label{eq:TvsR2}
{\nT} \approx 1-i{\nR}
\end{equation}
and
\begin{equation}
\label{eq:R_wRg0_wIl0} {\nR}(\omega) \approx -i
\frac{e^{2i\omega\delta}+1}{e^{4\pi\omega\sigma_+}+e^{2i\omega\delta}}
\fin
\end{equation}
Therefore,
\begin{equation}
\label{eq:T_wRg0_wIl0} {\nT}(\omega) \approx
\frac{e^{4\pi\omega\sigma_+}-1}{e^{4\pi\omega\sigma_+}+e^{2i\omega\delta}}
\fin
\end{equation}
Similarly, by the same method we used for $\omega_R<m\Omega$,
\begin{equation}
\label{eq:TR_wRl0_wIg0} {\nT}(-\omega) \approx 1 \lrgspc \mbox{and} \lrgspc
{\nR}(-\omega) \approx +i\fin
\end{equation}

\subsection{Results}
\label{subsec:Results}

It is convenient to introduce the notation
\begin{equation} \label{eq:SjDefinition}
S_j \equiv \int_{r_1}^{r_2} \omega V \, dr \coma
\end{equation}
where the subscript $j\in\{i,o\}$ indicates that the integration contour
crosses the real axis inside ($r_-<r<r_+$), outside ($r>r_+$) the event horizon.
Then
\begin{equation} \label{eq:SiSoSp}
i S_i = i \omega\delta \quad \mbox{and} \quad
iS_o=i\omega\delta-2\pi\omega\sigma_+\,.
\end{equation}
Analytic expressions for $S_j$ can be directly obtained
in terms of elliptic integrals.
We shall sometimes use the notations $l_i\equiv l_2$ and $l_o\equiv l_4$, so
$l_j$ may be taken as the integration contour of $S_j$.
Both $S_i$ and $S_o$ are
real in the highly damped limit, because in that limit $r_1$ and $r_2$ are
connected by anti-Stokes lines
on both sides of the horizon
\footnote{More precisely, this also requires $\re(\omega)=m\Omega$.}
and along these lines
$\im(\omega V\, dr)=0$ by definition. Note that
\begin{eqnarray}
e^{2i(S_i - S_o)} & = & e^{4\pi\omega\sigma_+} \nonumber \\
& = & \mp e^{(\omega-m\Omega)/T_H} \coma
\end{eqnarray}
where the upper (lower) sign corresponds to fermions (bosons), hereafter.

Our results for the highly damped regime may now be summarized as
\begin{eqnarray}
\label{eq:TGeneral}
{\nT}(\omega) & \approx & \frac{-e^{-2i\varepsilon(S_i-S_o)}+1}{e^{2i\varepsilon S_o}+1} \nonumber \\
& \approx & \frac{e^{\varepsilon(\omega-m\Omega)/T_H} \pm 1} {e^{2i\varepsilon S_o}+1}
e^{-\varepsilon(\omega-m\Omega)/T_H} \,; \nonumber \\
{\nT}(-\omega) & \approx & 1 \coma
\end{eqnarray}
and
\begin{equation} \label{eq:RGeneral}
{\nR}(\omega) \approx -i\varepsilon \frac{e^{-2i\varepsilon S_i}+1} {e^{-2i\varepsilon
S_o}+1}\,; \quad {\nR}(-\omega) \approx \varepsilon i \coma
\end{equation}
where we defined
\begin{equation}
\varepsilon \equiv \mbox{sign}(\omega_R-m\Omega) \fin
\end{equation}
These results reflect the expected branch cuts in $\nT$ and $\nR$ at $\omega_R = m \Omega$.
In the case of $\nT$ there is no cut for $\omega_I>0$; this is a
consequence of the fact that in this regime the boundary condition at the horizon
is uniquely defined without analytic continuation, as described in
the appendix of \cite{Neitzke:2003mz}.

Our results also imply
\begin{eqnarray} \label{eq:TT}
{\nT}(\omega)\tT(-\omega) & \approx &
\frac{-e^{-2i\varepsilon(S_i-S_o)}+1}{e^{2i\varepsilon S_o}+1} \\
& \approx & \frac{e^{\varepsilon(\omega-m\Omega)/T_H} \pm 1} {e^{2i\varepsilon S_o}+1}
e^{-\varepsilon(\omega-m\Omega)/T_H} \nonumber
\end{eqnarray}
and
\begin{equation} \label{eq:RR}
{\nR}(\omega)\tR(-\omega) \approx \frac{e^{-2i\varepsilon S_i}+1} {e^{-2i\varepsilon
S_o}+1} \coma
\end{equation}
which we will use in our discussion of the greybody factors in \S\ref{sec:DecayRate}.
As a consistency check,
note that these results satisfy the analytically continued flux
conservation relation, Eq.~(\ref{eq:ContinuedFluxConservation}).

\subsection{Boltzmann weights and resonances}
\label{subsec:BoltzmannFactors}

Both ${\nT}$ and ${\nR}$ given in Eqs.~\eqref{eq:TGeneral}-\eqref{eq:RGeneral} have a suggestive structure.
Beginning from Eq.~\eqref{eq:SjDefinition}, expanding $V$ and $r_i$ around large $|\omega|$ gives
\begin{equation} \label{eq:Sasymp}
S_j=\frac{\omega - \widetilde{\omega}_j}{4i T_j}+O(|\omega|^{-1}),
\end{equation}
where
\begin{equation}
\label{eq:DefinitionTj} \frac{1}{2 T_j} = 2i\int_{\widetilde{r}_1}^{\widetilde{r}_2}
\frac{\sqrt{q_0}}{\Delta} \, dr
\end{equation}
and
\begin{equation} \label{eq:DefinitionWj}
\frac{\widetilde{\omega}_j}{2 T_j} = -2i\int_{\widetilde{r}_1}^{\widetilde{r}_2}
\frac{q_1}{2\Delta\sqrt{q_0}} \, dr \fin
\end{equation}
Each term $e^{2i\varepsilon S_j}$ in
Eqs.~(\ref{eq:TGeneral}) and (\ref{eq:RGeneral}) thus becomes
$\exp{[\varepsilon(\omega-\widetilde{\omega}_j)/2 T_j]}$, and may be
interpreted as a Boltzmann weight corresponding to frequency $\omega$, temperature
$2 \varepsilon T_j$ and chemical potential $\widetilde{\omega}_j$.
(Alternatively, frequency $\omega/2$, temperature $\varepsilon T_j$ and chemical potential
$\widetilde{\omega}_j/2$.)
Moreover,
each $T_j$ is real, because $S_j$ is real to leading order. In
addition, from Eqs.~(\ref{eq:deltaDef}),(\ref{eq:SiSoSp}), $T_o < 0 \leq T_H/2 \leq T_i$, and
\begin{equation} \label{eq:TiToTH}
\frac{1}{2 T_i} - \frac{1}{2 T_o} = \frac{1}{T_H} \, .
\end{equation}
Similarly,
\begin{equation} \label{eq:WTiWTo}
\frac{\widetilde{\omega}_i}{2 T_i} - \frac{\widetilde{\omega}_o}{2 T_o} =
\frac{m\Omega}{T_H} + 2\pi i s \coma
\end{equation}
and $\re(\widetilde{\omega}_j)\propto m$ according to
Eq.~(\ref{eq:DefinitionWj}).

In our conventions, $T_o$ is negative and $T_i$ positive. However, as the
Boltzmann weights appear with different signs in
Eqs.~(\ref{eq:TGeneral}) and (\ref{eq:RGeneral}), the opposite convention would have
been equally natural.
We give a speculative thermodynamic interpretation of Eqs.~(\ref{eq:TiToTH}) and
(\ref{eq:WTiWTo}) in \S\ref{subsec:DecayRateInterpretation}.

\begin{figure}[h]
\centerline{\epsfxsize=7.5cm \epsfbox{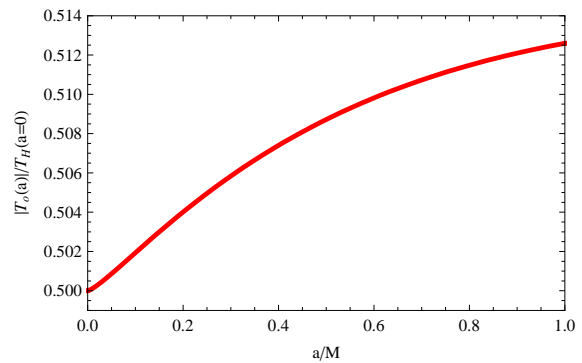}} \caption{\label{fig:To}
The effective temperature $|T_o|$ as a function of $a$ for an uncharged
rotating black hole, normalized by $T_H(a=Q=0)$.
}
\end{figure}

In Figure \ref{fig:To}, $|T_o|$ is plotted as a function of $a$ for $Q=0$, showing
that $T_o(a) \approx -T_H(a=0)/2$ within $\sim3\%$ accuracy.
Eq.~(\ref{eq:TiToTH}) then yields $T_i(a)^{-1} \approx
2[T_H(a)^{-1}-T_H(a=0)^{-1}]$ to this accuracy, so we do not plot $T_i$
independently.

Noting that ${\nT}$ and ${\nR}$
diverge when
\begin{equation} \label{eq:FirstQNMs}
\omega(n) = \widetilde{\omega}_o - 4 \pi i T_o(n+1/2)
\end{equation}
for integer
$n$, we may identify $4 \pi i T_o$ and $\widetilde{\omega}_o-2\pi i T_o$ respectively
as the level spacing and the offset of the highly damped QNM frequencies. For example, the real
part of the highly damped QNMs asymptotically approaches
$\re(\widetilde{\omega}_o)\propto m$. The QNM spectrum Eq.~(\ref{eq:FirstQNMs}) was shown in
\cite{Keshet:2007nv} to agree with previous numerical computations
\cite{Berti:2004um}.

In a similar fashion, $4\pi i T_i$ and $\widetilde{\omega}_i-2\pi i T_i$ are shown in
\S\ref{sec:Excitations} to be respectively the level spacing and offset
characterizing another type of resonant frequencies known
as total transmission modes (TTMs).
The asymptotic frequencies of the TTMs of a
rotating black hole have so far been unknown. Low-lying TTMs of a Schwarzschild
black hole were discussed in \cite{Andersson:1994tt,MaassenvandenBrink:2000ru}.
In the extremal limit $a \to M$ we have
$T_i \to 0$ and $\widetilde{\omega}_i \to m\Omega$,
so the TTMs coalesce to frequency $m\Omega$.
In the limit $a,Q \to 0$ we have $T_i \to \infty$, so no TTMs
exist in this limit in the highly damped regime.

\section{Resonances as excitations}
\label{sec:Excitations}

In this section we further examine the asymptotically damped black
hole resonances.  These resonances include the standard quasinormal modes (QNMs), but also include
other interesting families of modes; one might call all of them ``quasinormal''
in the sense that they decay with time, but in what follows we stick to the standard
terminology.

As we will see, each mode that we discuss can be associated with a semiclassical
state localized along one or two specific anti-Stokes lines,
independent of the boundary conditions at the horizon or spatial infinity.
The corresponding eigenstates and eigenvalues depend only on the integral of
the potential
$V$ along these lines. The eigenvalue frequencies of the various modes
satisfy a complex Bohr-Sommerfeld equation. In the QNM case, this equation
was shown in \cite{Keshet:2007nv} to reproduce earlier numerical results.
Our analysis suggests that in the highly-damped regime,
scattering off the black hole can be
effectively described in terms of a few coupled, one-dimensional,
semiclassical systems. This picture fully reproduces the resonances inferred from
\S\ref{sec:Analysis}.

This section is organized as follows.
In \S\ref{subsec:RealExcitations} we show that the wave equation becomes
semiclassical for the inverted potential which appears naturally along
anti-Stokes lines, define the corresponding eigenstates and derive
their eigenvalues.
Next, we discuss four types of eigenstates: (i) excitations along $l_4$
corresponding to quasinormal modes are discussed in \S\ref{subsec:QNMs};
(ii) excitations along $l_2$ corresponding to total transmission modes are
described in \S\ref{subsec:TTMs}; (iii) excitations circling $l_2$ and
$l_4$, corresponding to total reflection modes, are discussed in
\S\ref{subsec:TRMs}; and (iv) internal excitations along
$l_5$, associated with the behavior around $r_-$, are discussed in
\S\ref{subsec:IHMs}.  This last family of excitations does not appear directly
in $\nT$ or $\nR$, so they are not strictly speaking resonances of the black hole,
but from our present point of view they appear to be natural objects to consider.

The main properties of these modes are summarized in
Table \ref{tab:modes}. The emerging picture of a connected system of
black hole excitations is summarized in \S\ref{subsec:emerging_picture}.

\subsection{Highly-damped resonances as semiclassical excitations of the inverted potential}
\label{subsec:RealExcitations}

Eq.~(\ref{eq:WaveEquation}) can be interpreted as a Schr\"odinger equation
describing a particle of ``energy'' $E_z=\omega^2$ subject to a potential
$V_z$.
When $|\omega_I|$ is very large, $E_z$ is
approximately real and negative, so
we are looking at the
classically forbidden case $E_z \ll -|V_z|\leq 0$. However, the problem can be
continued to a classically-allowed one, by replacing $z$ with a Wick rotated
coordinate $x\equiv i z$, giving
\begin{equation} \label{eq:wave_x}
\left[-\frac{\partial^2}{\partial x^2}+(-V_z)-(-\omega^2)\right]f(x) = 0 \fin
\end{equation}
This is now a Schr\"odinger equation for a particle with energy $E_x=-\omega^2$
in the inverted potential $V_x(x)=-V_z$. The energy is approximately real and positive and $|V_x| \ll
E_x$ almost everywhere, motivating a semiclassical analysis. The coordinate $x$ is in
general complex, but it is approximately real along contours where $\re(\omega x)=0$.
These contours are the anti-Stokes lines defined by $\re(i\omega z)=0$,
discussed in \S\ref{sec:Analysis} and depicted as solid contours in Figure
\ref{fig:Stokes}. To avoid confusion, henceforth we refer to these
contours as excitation lines.

Although $V_x$ is in general complex, this makes little difference when
$|V_x|\ll E_x$, which holds true along most of each
excitation line $l$. This condition breaks down near the turning points
$x_i=i z_i$, but in these regions
\begin{equation} \label{eq:VxDivergence}
V_x(x\simeq x_i) \simeq -\frac{5}{36}(x-x_i)^{-2}
\end{equation}
is real and negative along $l$, so Eq.~(\ref{eq:wave_x}) can still be considered
as a real Schr\"odinger equation.
Furthermore, $V_x$ diverges at the turning
points, suggesting that the excitation lines can be regarded as
one-dimensional potential wells.
We may therefore study bound states, determined by applying
the wave equation~(\ref{eq:wave_x}) to each excitation line $l$ in the
system.
Note that
black hole QNMs have previously been studied by inverting the potential
and mapping the resonances to bound states, in special cases (for example scattering
off a slowly rotating black hole in the eikonal limit)
where the potential can be approximated by a
P\"{o}schl-Teller potential \cite{fermas}.

In the highly-damped limit, the eigenstates and eigenvalues corresponding
to the bound states are determined, as usual, by
a Bohr-Sommerfeld rule derived from
the semiclassical (WKB) approximation
\begin{eqnarray} \label{eq:BS_general_l}
\pi\left( n+\frac{\mu}{4} \right) & \approx & \int_l p_x \, dx \nonumber \\
& \approx & \int_{l} \sqrt{E_x-V_x}\,dx \nonumber \\
& \approx & \int_{l} \sqrt{\omega^2 - V_z}\,V \,dr \coma
\end{eqnarray}
where $p_x$ is the classical momentum corresponding to
Eq.~(\ref{eq:wave_x}),
and $n\in\Z$, where $\abs{n} \gg 0$ is the number of nodes of $f$ along $l$.
The number $\mu$ is the Maslov index \cite[see for example Ref.][]{tabor-ch}
which counts the $\pi/4$ phase shifts associated with the
turning points traversed by $l$. In the highly damped limit, to order
$O(|\omega|^{-1})$ Eq.~(\ref{eq:BS_general_l}) becomes
\begin{equation} \label{eq:BS_l}
S_j \equiv \int_{l_j} \omega V\,dr \approx \pi\left( n+\frac{\mu_j}{4} \right)
\coma
\end{equation}
where $j$ is the index of the excitation line or combination of lines.
With the appropriate choice of orientation for $l_j$, we may
identify the classical actions $S_2$ and $S_4$ with the $S_i$ and $S_o$ defined in
Eq.~(\ref{eq:SjDefinition}).

As in Eq.~\eqref{eq:Sasymp}, we expand $S_j= (4i T_j)^{-1}(\omega -
\widetilde{\omega}_j) + O(|\omega|^{-1})$, with $T_j$ and
$\widetilde{\omega}_j$ defined as in
Eqs.~(\ref{eq:DefinitionTj})-(\ref{eq:DefinitionWj}). This yields the
discrete, infinite eigenvalue spectrum of excitation frequencies
\begin{equation} \label{eq:resonances_l}
\omega_j(n) = \widetilde{\omega}_j + 4\pi i T_j \left( n+\frac{\mu_j}{4}
\right) \coma
\end{equation}
generalizing the QNM condition of Eq.~(\ref{eq:FirstQNMs}).
The resonances all have $\omega_I<0$ (recall that for $\omega_I>0$, $\nT$ and $\nR$ are constants),
so $n T_j<0$.
Recall that $S_j$ and $T_j$ are purely real in the highly damped limit
because along the excitation
lines, by definition, $\omega V\, dr\in\R$.
Eq.~\eqref{eq:DefinitionWj} implies that
$\re(\omega_j)=\re(\widetilde{\omega}_j)\propto m$; in particular, when $m=0$, the real parts of the
resonant frequencies vanish to order $|\omega|^{-1}$.

The presence of bound states in the system, if only along specific lines
in the complex $r$-plane, suggests that their eigenvalues may have physical
significance. Indeed, in \S\ref{subsec:QNMs}-\S\ref{subsec:IHMs} it is
shown that applying Eq.~(\ref{eq:BS_l}) to each excitation line reproduces
a certain resonance mode of the black hole. For example, excitations along
$l_4$ correspond to the QNMs. Note that this definition
of the excitations does not involve fixing the boundary condition at
spatial infinity or at the horizons. Rather, Eq.~(\ref{eq:BS_l})
determines the semiclassical eigenstates
locally, purely in terms of (the integral of) $V$ along $l_j$.

The Stokes phenomenon determines the relation between the wavefunctions along
adjacent excitation lines. This, as well as the exponential decay of the
wavefunction in time, makes it natural to view the excitation lines as coupled to
one another.
Indeed, the analysis of the transmission-reflection problem in \S\ref{sec:Analysis}
could be rephrased in the language of tunneling through the potential barriers at the
turning points; we discuss this in \S\ref{subsec:emerging_picture}.

For convenience we
define $\rho \equiv -i\omega$, such that WKB modes
$f_\pm = e^{\pm i \omega z} = e^{\pm i \rho x}$ with a plus (minus)
sign travel toward (away from) spatial infinity.

\subsection{Quasinormal modes}
\label{subsec:QNMs}

The most familiar type of black hole resonance is a quasinormal mode
(QNM). These linear, damped
modes dominate the intermediate-time behavior of black hole perturbations.
The discrete QNM frequencies, which correspond to poles of the
transmission and reflection amplitudes ${\nT}$ and ${\nR}$, may be determined by
studying perturbations that satisfy purely outgoing boundary
conditions at both the event horizon and spatial infinity along the
physical interval $r_+<r<\infty$.
The highly-damped QNM frequencies were derived analytically for
spherically-symmetric black holes in \cite{Motl:2003cd}, and for a
rotating black hole in \cite{Keshet:2007nv}.

Now we propose to identify these resonances with bound states confined
along an excitation line.  Which line should we consider? In the classical
picture of the QNM, the potential barrier on the interval $r_+ < r <
\infty$ plays an important role; one pictures this barrier as ``ringing''
and emitting energy toward the horizon and spatial infinity. This
motivates the suggestion that the QNMs correspond to the excitation line
$l_4$, as it intersects the real $r$-axis at a point $r_{l4}$ located just
outside the event horizon.  A
second motivation is that ${\nR}$ and ${\nT}$, both of which
develop a pole at the QNM frequencies, are the amplitudes of the WKB modes
$f_\pm\propto e^{\pm i \rho x}$ along $l_4$ (see Figure \ref{fig:Computation}).
These rough arguments lead to the right conclusion: Eq.~(\ref{eq:BS_l})
applied to $l_4$, with $\mu=2$ phase shifts associated with $r_1$ and
$r_2$, precisely agrees with the highly damped QNM condition of
\cite{Keshet:2007nv} for a rotating black hole. This equation may be
rewritten as
\begin{equation}
\label{eq:QNM_condition_l4} e^{2iS_4}+1 = \exp{\left(2i\int_{l_4} \omega
V\, dr\right)} + 1 = 0 \coma
\end{equation}
which is indeed the location of the poles in ${\nT}$ and ${\nR}$, as seen from
Eqs.~(\ref{eq:TGeneral})-(\ref{eq:RGeneral}).

As $\omega$ approaches one of the QNM frequencies given by Eq.~(\ref{eq:QNM_condition_l4}),
${\nT}$ and ${\nR}$ diverge while satisfying $|{\nT}| \approx |{\nR}|$.
So the QNM excitations reduce to standing waves along $l_4$, decaying
exponentially in time.  One might heuristically understand this time decay
as follows:
for $\omega_R<m\Omega$ ($\omega_R>m\Omega$),
the outgoing -- into the black hole -- part of the
wavefunction, ${\nT}e^{-i\rho x}$, gradually tunnels across the turning point
into $l_2$ [and into $l_3$ ($l_1$)], whereas the ingoing part, ${\nR}e^{+i\rho x}$, tunnels its way
to $l_1$ ($l_3$), thereafter
escaping to spatial infinity.

\begin{table*}
\caption{\label{tab:modes} Highly damped resonances of a rotating black
hole. }
\begin{ruledtabular}
\begin{tabular}{ccccccc}
Mode & $f(r_+)$\footnotemark[1] & $f(r\rightarrow\infty)$\footnotemark[1]
& Excitation line & Excitation mode & Eigenvalue equation & Section \\
\hline
QNM & $f_-\uparrow$ & $f_+\uparrow$ & $l_4$ & $f_+-i\varepsilon f_-$ & $e^{2iS_4}+1=0$ & \ref{subsec:QNMs} \\
TTM & $f_-\uparrow$ & $f_-\downarrow$ & $l_2$ & $f_++i\varepsilon f_-$ & $e^{2iS_2}+1=0$ & \ref{subsec:TTMs} \\
TRM & $f_+ = f_-$ & $f_+-i \varepsilon f_-$ & $l_2$ and $l_4$ &
$-i\varepsilon f_+$ & $e^{2i(S_2-S_4)}-1=0$ & \ref{subsec:TRMs} \\
IHM & & & $l_5$ & $f_+$ & $e^{2iS_5}-1=0$ &
\ref{subsec:IHMs}
\end{tabular}
\end{ruledtabular}
\footnotetext[1]{Arrows pointing up (down) indicate diverging (vanishing) $f_\pm$. }
\end{table*}

\subsection{Total-transmission modes}
\label{subsec:TTMs}

A less frequently explored type of black hole resonance is the
total-transmission mode (TTM) \footnote{See for example \cite{Leung:1999fr}.
TTMs are also known as transmission resonances \cite{bohm-qt} or left mixed
modes \cite{Fiziev:2005ki}.}. These modes, defined by ${\nR}=0$,
can be studied as perturbations that are purely outgoing at the event horizon and
purely \emph{ingoing} at spatial infinity.

Like the QNMs, the TTMs are associated with a specific excitation line. To
guess which line it should be, note that Eqs.~(\ref{eq:TvsR}), (\ref{eq:TvsR2})
give ${\nT}\approx 1$.  This implies that the
wavefunction along $l_2$ becomes a (damped) standing wave, $f(l_2)\approx
-i\varepsilon e^{+i\rho x} + e^{-i\rho x}$, suggesting that excitations along this
line could correspond to the TTMs. Furthermore, along $l_2$ the reflection
amplitude ${\nR}$ does not appear as the coefficient of either WKB
component (see Figure \ref{fig:Computation}).
Indeed, applying Eq.~(\ref{eq:BS_l}) to $l_2$ yields
\begin{equation}
e^{2iS_2} + 1 = \label{eq:TTM_condition_l2} \exp{\left(2i\int_{l_2} \omega
V\, dr\right)} + 1 = 0 \coma
\end{equation}
which is the condition for the numerator of ${\nR}$ to vanish,
thus determining the TTM frequencies.
Note that $f(l_2)\approx -i\varepsilon e^{+i\rho x} + e^{-i\rho x}$ implies
that $c_+(l_1,l_3)=0$ for $\omega_R<m\Omega$, $\omega_R>m\Omega$, so the TTM excitation cannot escape
from $l_2$ to spatial infinity.

Total transmission modes occur in various physical settings in
which two systems are connected through tunneling across a barrier.
It is generally found that the frequencies of total transmission into
a system coincide with its metastable eigenfrequencies \cite{bohm-qt}.
This suggests that the TTM frequencies of a
black hole could coincide with the eigenenergies of some internal black hole
degrees of freedom. In a sense this is what we have found in the highly damped
limit:  the TTM frequencies of the classical black hole coincide with the energies
of bound states along the line $l_2$, which is ``internal'' to the black hole
in the sense that it meets the real axis at a point $r_{l2}$ behind
the event horizon, $r_-<r_{l2}<r_+$.

The description of the TTMs as excitations along $l_2$ uses the
analytic continuation of the metric behind the
event horizon.  The physical significance of such a continuation is of course
unclear.  However, we emphasize that the resonant frequencies themselves do not
depend on this continuation. The modes may be defined by imposing the
appropriate boundary conditions at $r_+$ and as $r\rightarrow\infty$. The
resonant frequencies can then be derived using Teukolsky's equation along $r_+<r<\infty$,
for example in the method of \cite{leaverkerr}.

\subsection{Total-reflection modes}
\label{subsec:TRMs}

Black holes also have modes of total reflection, where ${\nT} = 0$
\footnote{These total-reflection modes, also known as reflection
resonances, often occur when a metastable state destructively
interferes with the transmitted wave.}.
Using Eqs.~\eqref{eq:TvsR} and \eqref{eq:TvsR2}, these modes correspond to
standing wave behavior at spatial infinity, $f(r\rightarrow\infty)\propto
-\varepsilon ie^{+i \rho x}+e^{-i \rho x}$,
and equivalently along $l_1$ ($l_3$) for $\omega_R<m\Omega$ ($\omega_R>m\Omega$).

When ${\nT}\approx 0$, the wavefunction assumes the same form along $l_2$ and
along $l_4$, $f\propto e^{+i\rho x}$, describing a purely traveling wave.
The TRMs can therefore be identified for $\omega_R<m\Omega$ as excitations clockwise circling
$l_2$ and $l_4$, traveling from $r_1$ to $r_2$ along $l_2$ and back to
$r_2$ along $l_4$, and vice versa for $\omega_R>m\Omega$. These
modes travel in a closed loop unaffected by the turning points ($f_+$ is
subdominant within the loop), implying a Maslov index $\mu=0$. Hence
applying Eq.~(\ref{eq:BS_l}) to the $l_2-l_4$ contour yields
\begin{eqnarray}
\label{eq:TRM_condition_l2l4} e^{2i(S_2-S_4)}-1
& = & e ^{4\pi \omega \sigma_+} -1 = 0 \fin
\end{eqnarray}
This result is the condition for the numerator of ${\nT}$ to vanish in
Eq.~(\ref{eq:TGeneral}), and
therefore indeed determines the TRM frequencies. Note that modes with the
opposite orientation, counterclockwise (clockwise) rotating for
$\omega_R<m\Omega$ ($\omega_R>m\Omega$), are precluded by the Stokes phenomenon (such
a mode would be dominant on the Stokes lines which run
to $r_+$, but then crossing these lines would introduce components of the
other WKB mode).

The integral in Eq. \eqref{eq:TRM_condition_l2l4} can be evaluated by residues, in which case
the only contribution comes from the singularity at
$r_+$.  This suggests that the TRMs are in some sense associated with the event horizon.
Note also that the expression $\mp(e^{4\pi\omega\sigma_+}-1) =
e^{(\omega-m\Omega)/T_H}\pm 1$
is the inverse of the spectrum of Hawking's thermal
radiation from the horizon.
The association between TRMs and Hawking radiation will be revisited
in \S\ref{sec:DecayRate}.

The TRM frequencies inferred
from Eq.~(\ref{eq:TRM_condition_l2l4}) are
\begin{equation} \label{eq:TRM_freq}
\omega_{TRM}(m) = m\Omega - 2\pi i T_H \left(n-s \right) \fin
\end{equation}
This expression for the TRM frequencies holds also for non-rotating black
holes, where $\Omega=0$.  In \S\ref{subsec:HawkingCancellation} it is shown
that Eq.~(\ref{eq:TRM_freq}) is exact --- there are no $O(|\omega|^{-1})$
corrections.

\subsection{Inner horizon modes}
\label{subsec:IHMs}

One more excitation line, $l_5$, lies in the $\re(r)>0$ region.
This line emanates from the turning
point $r_0$ and circles the inner horizon $r_-$, as shown in
Figure \ref{fig:Stokes}
\footnote{This line is shown for $\omega_R=m\Omega_-$. We ignore an additional finite excitation
line asymptotically connecting $\widetilde{r}_0=0$ and $\widetilde{r}_3<0$ (for $Q=0$).}.
Excitations associated with $l_5$ are not directly
relevant to the scattering process discussed in \S\ref{sec:Analysis} and do not appear in $\nT$ and $\nR$,
because this line is not directly connected to the lines $l_1-l_4$. We may
nevertheless calculate the eigenstates and eigenvalues of excitations
associated with $l_5$. The excitation frequencies are given by
Eq.~(\ref{eq:BS_l}), with integration carried out along $l_5$ and $\mu=0$.
The only contribution to the integral arises from the singularity at
$r_-$. The result is
\begin{eqnarray}
\label{eq:IHM_condition_l5} 2\pi n & = & 4\pi i \omega
\underset{\,\,\,r\rightarrow
r_-}{\mbox{Res}} (V) \\
& = & i \frac{\omega-m\Omega_-}{T_-} + 2\pi s \nonumber \coma
\end{eqnarray}
where $T_-=-(r_+-r_-)/A_-<0$
and $\Omega_-\equiv 4\pi a/A_-$ are the temperature and angular velocity of
the inner horizon, respectively. As in the case of TRMs, only one
orientation, $f\propto e^{+i\rho x}$, is possible.

Eq.~(\ref{eq:IHM_condition_l5}) and the resonant frequencies it implies,
\begin{equation}
\omega = m\Omega_- - 2\pi iT_- \left( n+s \right) \coma
\end{equation}
demonstrate that these modes are associated with the inner
horizon. The excitation line $l_5$ does cross the real axis near $r_-$, at
two points: close to the ring singularity $\widetilde{r}_0=0$ (if $Q=0$)
and at a point $r_{l5}$ lying
between $r_-$ and $r_{l2}$, so $r_-<r_{l5}<r_{l2}<r_+$. We therefore call
these modes inner horizon modes (IHMs). There is a formal resemblance between
the IHMs and the TRMs, the latter similarly associated with the outer horizon.

Although $l_5$ is not connected to
the other excitation
lines discussed above, there is a special case where we can nevertheless
relate $l_5$ to the boundary condition at spatial infinity. Namely, when
the latter is purely outgoing [$f(r\rightarrow\infty)\propto e^{i\omega
z}$], the asymptotics at $l_2$ can be continued directly to $l_5$, implying
that $f(l_5)\propto e^{-i\omega z}\propto e^{-i\rho x}$.
Such a continuation cannot be carried out for more general
boundary conditions at spatial infinity.

\subsection{Summary: connected semiclassical systems}
\label{subsec:emerging_picture}

The results of this section show that highly-damped
perturbations of a rotating black hole may be described in
terms of three inter-connected lines: (i) $l_1$ or $l_3$ (depending on $\varepsilon$),
admitting waves that travel to/from spatial infinity;
(ii) $l_4$, corresponding to the near environment of the black hole,
carrying the QNM excitations that can tunnel out to $l_2$ and to infinity
through $l_1/l_3$; and (iii) $l_2$, describing some internal black hole
region between $r_-$ and $r_+$ and carrying the TTM excitations, which can
be excited by a wave incident from spatial infinity but cannot directly escape to
infinity.
Combined, $l_2$ and $l_4$ form a loop that carries the TRM excitations,
modes circling the event horizon which are related to Hawking radiation.
Each Boltzmann factor (see
\S\ref{subsec:BoltzmannFactors}) in Eqs.~(\ref{eq:TT}) and (\ref{eq:RR})
is associated to one of these types of excitations.

Each of the excitation lines is connected to two other lines at the
turning points. Since the effective potential diverges at these turning
points, we can view each excitation line as a ``potential well''
supporting bound states.
The wavefunction can tunnel from one line to an adjacent one while picking up
a phase shift, as dictated by the Stokes phenomenon.
This provides a heuristic picture of the manner in which excitations can
decay and possibly interact.

Each excitation line $l_j$ crosses the real axis at a single point
$r_{lj}$, corresponding physically to an equatorial ring
(\S\ref{subsec:EquatorialFocusing}); $l_4$ corresponds to a ring just
outside the outer horizon, near the peak of the potential barrier, while
$l_2$ is associated with an internal ring lying between $r_-$ and $r_+$.
The complex-plane connections between the different excitation lines
directly relate the behavior of the perturbation along disconnected,
distant rings.

\section{Complex Geodesics}
\label{sec:Geodesics}

In the preceding sections, the one-dimensional wave equations
(\ref{eq:WaveEquation}) and (\ref{eq:wave_x}) were analyzed with little reference to the
underlying (3+1)-dimensional metric.  Since radiation propagates
along null geodesics in the large $\omega$ limit, one might expect
that quantities playing a role in our analysis,
such as the characteristic spacing and offset of
the resonant frequencies, should be understandable in terms of
null geodesics in the complexified metric.  In this section we show
that this is indeed the case.

In \S\ref{subsec:GeneralGeodesics} we review some generalities on the
analytically continued null geodesics and identify $r_{1,2}$
as turning points of these geodesics in the small impact parameter limit.
In \S\ref{subsec:EquatorialGeodesics} we
focus our attention on geodesics in the equatorial plane, and show the
role they play in our analysis.

\subsection{Geodesics}
\label{subsec:GeneralGeodesics}

We study the complexified
geodesic trajectories of a massless particle with angular momentum $p_\phi=m$,
complex energy $E=\omega$, and Carter's (fourth) constant of motion \cite{Carter:1968rr}
fixed to some $Q_C$.

Along a null geodesic, the derivatives of Boyer-Lindquist coordinates with
respect to the affine parameter $\lambda$ are then \cite{nov-frol}
\begin{equation} \label{eq:Geodesic_r}
\dot{r} = \frac{\sqrt{E^2 q_0 - 2 a (2 M r - Q^2) p_\phi E -
(\Delta-a^2)p_\phi^2-\Delta Q_C}}{\Sigma} \coma
\end{equation}
\begin{equation} \label{eq:Geodesic_t}
\dot{t} = \frac{[(r^2+a^2)^2-a^2 \Delta \sin^2\theta] E - a (2 M r -Q^2)
p_\phi} {\Sigma\Delta} \coma
\end{equation}
\begin{equation} \label{eq:Geodesic_phi}
\dot{\phi} = \frac{a(2 M r -Q^2) E+(\Delta \sin^{-2}\theta -a^2)p_\phi} {
\Sigma\Delta} \coma
\end{equation}
and
\begin{equation} \label{eq:Geodesic_theta}
\dot{\theta} = \frac{\sqrt{ a^2 E^2 \cos^2\theta - p_\phi^2
\cot^2\theta+Q_C}} {\Sigma} \coma
\end{equation}
where the square root branches in Eqs.~(\ref{eq:Geodesic_r}) and
(\ref{eq:Geodesic_theta}) are chosen independently,
and we recall $\Sigma=r^2+a^2 \cos^2 \theta$.
To leading order in $|\omega|^{-1}$,
Eq.~(\ref{eq:Geodesic_r}) becomes $\dot{r}\approx Er^{-2}q_0^{1/2}$; so
in the highly damped limit $r_1$ and $r_2$ approach the turning
points of the complexified geodesics where $\dot{r}=0$.

The covariant momentum $p_r$ is determined by the constants of motion as
\cite{Bekenstein:1973ur}
\begin{eqnarray}
(p_r \Delta)^2 & = & q_0 E^2 -
2a(2Mr-Q^2)Ep_\phi \nonumber \\
& &   - (\Delta-a^2)p_\phi^2 - Q_C\Delta \fin
\end{eqnarray}
Using this together with Eqs.~(\ref{eq:V_definition})-(\ref{eq:q1_definition}), the quantity
$\omega V$ which was crucial for the WKB analysis may be expanded around large $\omega$ as
\begin{equation} \label{eq:QuantizedAction}
\omega V = p_r+i s V_s+i A_1 V_A + O(|\omega|^{-1}) \coma
\end{equation}
where we defined
$V_s \equiv q_0^{-1/2}\Delta^{-1}[r(\Delta+Q^2)-M(r^2-a^2)]$ and $V_A \equiv -
q_0^{-1/2}a/2$.
The resonant frequency equation ~\eqref{eq:BS_l} can now be written to
order $|\omega|^{0}$ as a complexified Bohr-Sommerfeld rule \cite{Keshet:2007nv}
\begin{equation} \label{eq:QuantizedZeroAction}
2\int_{l_j} p_r \, dr =
\pi \left(n+\frac{\mu_j}{4}\right) \coma
\end{equation}
where the excitation lines $l_j$ can be understood as contours of steepest
descent of $\omega V$ connecting the geodesic turning points $r_i$ which lie
at the endpoints of $l_j$. In order to reproduce
the resonant frequencies to order $|\omega|^{-1}$, the integrand should be
replaced by $\widetilde{p}_r=p_r+i s V_s+i A_1 V_A$.

\subsection{Equatorial geodesics}
\label{subsec:EquatorialGeodesics}

Focusing on the equatorial region, we may replace
Eqs.~(\ref{eq:Geodesic_r})-(\ref{eq:Geodesic_theta}) by the lowest order
terms in their expansion about $\theta=\pi/2$. To this order, $Q_C=0$. On
the equator, $\dot{t}$ also vanishes to leading order in $|\omega|^{-1}$
at the turning points $r_i$, regardless of $\arg(\omega)$.
More generally, on the equator $r_1$ and $r_2$ are turning
points where both $\dot{r}$ and $\dot{t}$ vanish simultaneously,
in the limit of small impact parameter $b\equiv p_\phi/E$ in which $|b| \ll a$,
and in particular when $p_\phi=0$.

The Boltzmann factors of \S\ref{subsec:BoltzmannFactors} can now be
related to the equatorial null geodesics. Consider the
expansion of the action $S \approx (\omega-\widetilde{\omega})/(4iT)$,
where $T$ and $\widetilde{\omega}$ are given by
Eqs.~(\ref{eq:DefinitionTj})-(\ref{eq:DefinitionWj}) and we have omitted
the index $j$ of the excitation lines for brevity. A direct comparison between these
quantities and Eqs.~\eqref{eq:Geodesic_r}-\eqref{eq:Geodesic_phi}, after
substituting $\theta = \pi/2$, gives
to leading order in $|\omega|^{-1}$
\begin{equation} \label{eq:TusingGeodesics}
\frac{1}{T} \approx 4i\int \frac{dt}{dr} dr = 4i\Delta t
\end{equation}
and
\begin{equation} \label{eq:WjIsDphi}
\re(\widetilde{\omega}) \approx 4iT m \int \frac{d\phi}{dr} dr = m
\frac{\Delta \phi}{\Delta t} \coma
\end{equation}
where $\Delta t$ and $\Delta \phi$ are respectively the time and the
azimuthal angle elapsed along the integrated geodesic. Moreover, using
$V_A=-(2\cos\theta)^{-1}(\pm d\theta/dr)$,
\begin{equation} \label{eq:resonanceGeodesic}
\im(\widetilde{\omega}) \approx  \pm i A_1 \frac{\Delta \zeta}{\Delta t} -
i s \frac{\int  V_s \,dr}{\Delta t} \coma
\end{equation}
where we defined a logarithmically-stretched angular coordinate
\begin{equation} \label{eq:zetaDefinition}
\zeta (\theta) \equiv \int (2\cos\theta)^{-1} d\theta
\simeq \frac{1}{2}\ln\left(\theta-\frac{\pi}{2} \right) + \mbox{const} \coma
\end{equation}
the last approximation valid near $\theta = \pi/2$.

So the integral of $\omega V$ between any two
values of $r$ is
\begin{equation} \label{eq:GeodesicAction}
S = \omega \Delta t - m \Delta \phi \mp i A_1 \Delta \zeta  + i s \int V_s
\,dr +O(|\omega|^{-1}) \fin
\end{equation}
The solution to the transmission-reflection problem in
Eqs.~(\ref{eq:TT})-(\ref{eq:RR}) may thus be rewritten in terms of the
more physical quantities associated with a null geodesic.
Eq.~(\ref{eq:GeodesicAction}) is seen to be a restatement of the result
$S\approx \int \widetilde{p}_r \,dr$,
because along null geodesics
$p_r\,dr = \omega\,dt - m\,d\phi - p_\theta\, d\theta$.

It follows that the highly-damped resonant frequencies corresponding to a given excitation
contour $l$ are determined by
\begin{eqnarray} \label{eq:WnUsingGeodesics}
\omega(n)\Delta t & = & m \Delta \phi \pm i A_1 \Delta \zeta - i s
\int_l V_s \,dr \nonumber \\
& & + \pi\left( n+\frac{\mu}{4} \right) \coma
\end{eqnarray}
where $\Delta t$, $\Delta \phi$, $\Delta \zeta$, and $\int V_s \,dr$ are
calculated along $l$, and are all imaginary. As an example, for a closed,
clockwise contour that encircles $r_+$ we find $\Delta t=(2iT_H)^{-1}$,
$\Delta \phi=\Omega\Delta t$, $\Delta \zeta = 0$, and $\int V_s\,dr=i\pi$.
Plugging these quantities into Eq.~(\ref{eq:WnUsingGeodesics}) with
$\mu=0$ yields the TRM frequencies of Eq.~(\ref{eq:TRM_freq}).

Altogether we have found that the resonant frequencies $\omega(n)/2\pi$ can be
understood as harmonics of a fundamental (imaginary)
frequency $(2\Delta t)^{-1}$ plus an offset $\widetilde{\omega}/2\pi+\mu/8\Delta t$,
such that $\Delta t$ and $\widetilde{\omega}\Delta t$ are associated
respectively with the time and with a generalized angular distance
(including $m \Delta \phi$ and $i A_1 \Delta \zeta$, as well as $\mu$- and spin terms) elapsed
along a null geodesic corresponding to the relevant excitation line.
Somewhat similar connections have been suggested by studies of black holes in the
eikonal limit, where approximate expressions for the QNMs were inferred
from the decay of wavepackets which travel initially along unstable closed orbits
\cite{goebel-qnm, fermas}.

\section{Black hole decay and greybody factors}
\label{sec:DecayRate}

In this section we sift the results of the preceding sections for clues
about the quantum description of the black hole spacetime. The
analytically continued spectrum of Hawking radiation escaping from the
black hole is presented in \S\ref{subsec:DecayRate} and
\S\ref{subsec:HawkingCancellation}. In
\S\ref{subsec:DecayRateInterpretation} we recall some examples where a similar
spectrum was found to correspond to a dual conformal field theory (CFT),
and speculate on the microscopic description underlying the present case.

\subsection{Decay spectrum}
\label{subsec:DecayRate}

First, recall that for real frequency $\omega$ the transmission amplitude provides
information about the Hawking radiation emitted from the black hole, as
observed from spatial infinity.  In \cite{Hawking:1975sw} it is argued that this observed
spectrum $\Gamma(\omega)$ is related to the absorption probability $\sigma(\omega)$ by
\begin{equation} \label{eq:emission}
\Gamma  = \frac{d^2 N}{dt\, d\omega} = \sigma(\omega) n_H(\omega) \coma
\end{equation}
where $n_H(\omega)$ denotes the spectrum of pure blackbody radiation at temperature $T_H$
and potential $m\Omega$,
and $\sigma(\omega)$ acts as a ``greybody factor'' which filters this thermal spectrum.
There is some arbitrariness in how one continues
Hawking's formula to complex $\omega$; we make a choice which will be convenient for
what follows, namely
\begin{equation} \label{eq:HawkingSpectrum}
n_H(\omega)=\frac{1}{e^{\varepsilon(\omega-m\Omega)/T_H}\pm1} \fin
\end{equation}
Upper (lower) signs correspond to emission of fermions (bosons), above and henceforth.

In \S\ref{subsec:TRProblem} we argued that $\sigma(\omega) =
{\nT}(\omega) \tT(-\omega)$. Now we analytically continue to the highly damped
regime. Using Eq.~(\ref{eq:TT}) then gives
\begin{align} \label{eq:DecayRate}
\Gamma(\omega)
& \approx \frac{e^{-\varepsilon(\omega-m\Omega)/T_H}}
{e^{\varepsilon(\omega-\widetilde{\omega}_o)/2 T_o}+1} \fin
\end{align}

\subsection{Exact cancellation of Hawking spectrum}
\label{subsec:HawkingCancellation}

In the expression for the decay spectrum in Eq.~(\ref{eq:DecayRate})
the pole of the spectrum $n_H$ in Eq.~(\ref{eq:HawkingSpectrum}) cancels with the zero of
${\nT}(\omega)\tT(-\omega)$ in Eq.~(\ref{eq:TT}). Based on our arguments so
far, though, one might have thought that this cancellation is only
approximate and the exact analytically continued spectrum would have poles
and zeroes separated by a distance $O(\abs{\omega}^{-1})$.

Actually, the zeroes and poles cancel one another exactly. The reason is
that the boundary condition Eq. \eqref{eq:boundary_conditions} manifestly
requires ${\nT}(\omega) \neq 0$, so ${\nT}(\omega) = 0$ is possible only if
Eq.~\eqref{eq:boundary_conditions} breaks down. But this equation breaks
down only when the two solutions near $r = r_+$ have the same monodromy,
since then we cannot pick out a solution uniquely by specifying its
monodromy. Inspection of Eqs.~(\ref{eq:BoundaryConditionsR}) or
(\ref{eq:SigmaPlus}) indicates that this condition is equivalent to
vanishing of the denominator of $n_H$ in Eq.~(\ref{eq:HawkingSpectrum}).
This argument applies quite generally, in particular to the spherical
black holes analyzed in \cite{Neitzke:2003mz}.

We have thus shown that ${\nT}$ can have zeros only where $n_H$ has poles.
This directly relates the TRM frequencies to the poles of $n_H$.
In Appendix \ref{subsec:GeneralizedTR} it is shown that ${\nT}$ does indeed have such
zeros in a large class of black holes in the highly-damped limit.

\subsection{Speculations on the microscopic description}
\label{subsec:DecayRateInterpretation}

As shown in \S\ref{subsec:DecayRate}, there is a pleasantly simple
expression for the decay spectrum at large imaginary frequencies, given in
Eq.~(\ref{eq:DecayRate}). But what could its physical meaning be?

\subsubsection{Examples of known dual CFTs}
\label{subsubsec:DualExamples}

Recall that computations of the same quantity at small real frequencies
have in the past given information about quantum gravity in black hole backgrounds
\cite{Maldacena:1997ix, Maldacena:1997ih, Aharony:1999ti}.  For example,
consider scalar emission from a four-dimensional, slowly rotating
($\Omega\ll 1/M$) black hole in the regime $\omega\ll 1/M$.
The corresponding decay spectrum given in \cite{Maldacena:1997ih} can be written as
\begin{equation} \label{eq:DecayBPS}
\Gamma(\omega) \propto \frac{\omega^{2l-1} P_{2l+1}(\omega)}{e^{(\omega-m\Omega)/T_H} -
1} \coma
\end{equation}
where $P_{2l+1}$ is a polynomial of order $2l+1$. Near BPS saturation
($Q=M-\epsilon$ and $a^2\sim Q\epsilon$ for small $\epsilon>0$)
the degrees of freedom of the black hole are described by a
chiral
$(0,4)$ superconformal
field theory, and a SCFT computation of the decay spectrum
agrees precisely with Eq.~(\ref{eq:DecayBPS}) \cite{Maldacena:1997ih}.

A second example is scalar emission from a five-dimensional, non-rotating black hole. In a
certain ``dilute gas'' limit, the decay spectrum is \cite{Maldacena:1997ih}
\begin{equation} \label{eq:5DDecay}
\Gamma(\omega) \propto \frac{\omega^{2l} P_{2l+1}(\omega)} {\left( e^{\omega/2T_L}
\pm 1 \right)\left( e^{\omega/2T_R} \pm 1 \right)} \coma
\end{equation}
where a positive (negative) sign corresponds to odd (even) $l$.
Again, this agrees with a stringy computation of the black hole decay spectrum \cite{Maldacena:1997ih}; these results
were important precursors of the AdS/CFT correspondence.

In both Eqs. \eqref{eq:DecayBPS} and \eqref{eq:5DDecay} there are characteristic
denominator factors, which have the form of partition functions of ensembles
constructed from the degrees of freedom of the microscopic CFT.  In the case
of Eq. \eqref{eq:DecayBPS} the relevant CFT is chiral, so we see only one type of
bosonic excitation, at temperature $T_H$.  In Eq. \eqref{eq:5DDecay}
the CFT is non-chiral, and the left-moving and right-moving sectors have
different temperatures $T_L$, $T_R$, obeying
\begin{equation} \label{eq:TLReqTH}
\frac{1}{2T_L}+\frac{1}{2T_R}=\frac{1}{T_H} \fin
\end{equation}
The appearance of a product of two denominator factors reflects the fact that
emission takes place only when left-moving and right-moving excitations collide.
Although the excitations can be fermionic or bosonic with conformal weights
$h_L=h_R=(l+2)/2$, bosonic statistics of the outcoming scalar emission is ensured by
$h_L-h_R=0$.

A third and last example is the
$(2+1)$-dimensional asymptotically anti-de Sitter BTZ black hole
\cite{Banados:1992wn}.  Here the QNM spectrum is given by
\cite{Birmingham:2001pj,Birmingham:2003wa},
\begin{equation} \label{eq:BTZ_QNMs}
\omega_{L,R} = k_{L,R}\sqrt{-\Lambda} -4\pi i T_{L,R}(n+h_{L,R}) \coma
\end{equation}
where $n, k_{L,R} \in\Z$, and $\Lambda$ is the cosmological constant.
The excitation temperatures $T_{L,R}$ characterize respectively the
left- and right-moving Virasoro algebras.
These temperatures also satisfy Eq.~(\ref{eq:TLReqTH}).
The angular momentum of the perturbation is given by
\begin{equation} \label{eq:BTZ_momentum}
k_L - k_R = \Delta J \fin
\end{equation}
The conformal weights $h_{L, R}$ satisfy
\begin{equation}
\label{eq:ConformalWeights}
h_L-h_R=\pm s \coma
\end{equation}
ensuring that the emitted Hawking quanta have the correct spin.
Unlike the previous examples, the Boltzmann factors here
involve chemical potentials with nonzero real part.

Note that Eqs.~(\ref{eq:TLReqTH}) and (\ref{eq:TiToTH}) are formally
identical (except for a sign in front of $T_o$, but recall we have chosen $T_o<0$). Similarly,
Eqs.~(\ref{eq:BTZ_momentum})-(\ref{eq:ConformalWeights}) are formally
identical to Eq.~(\ref{eq:WTiWTo}), if we define complex chemical potentials
$\widetilde{\omega}_{L,R}$ by
\begin{equation} \label{eq:LRPotentials}
\frac{\widetilde{\omega}_{L,R}}{2T_{L,R}} \equiv
\frac{\Omega}{T_H} k_{L,R} + 2\pi i h_{L,R} \coma
\end{equation}
with $\Delta J=m$ in the present study.

The decay spectrum of Eq.~(\ref{eq:DecayRate}) in the present analysis
contains a structure similar to the above examples: in particular a
Boltzmann weight with characteristic temperature and chemical potential
appears in the denominator, related to the highly-damped QNM spectrum. To
compare our results with the case of
a slowly rotating black hole in Eq.~(\ref{eq:DecayBPS}), consider the highly damped
results in the $a\to 0$ limit. Here $|2T_o|\to T_H$,
so the decay spectra in Eqs.~(\ref{eq:DecayRate}),(\ref{eq:DecayBPS}) have a similar
Boltzmann factor.
At low frequencies and non-negligible
rotation, the Kerr decay spectrum is probably more formally similar to the two
other (BTZ and extremal 5D) examples given above, because Kerr QNMs in
this regime fall into two families \cite{leaverkerr}, implying that two
Boltzmann-like factors appear in the denominator of $\Gamma$.

\subsubsection{Speculations}
\label{subsubsec:DualSpeculations}

By analogy with the cases just reviewed, we would like to interpret the decay spectrum we computed as
giving information about the microscopic degrees of freedom of the rotating black hole
in the highly damped frequency regime.  Here we present a few speculations in that direction.

We took $\abs{\omega}$ much larger than
all other scales, so one might expect that the physics in this regime is
scale invariant; hence we might try to interpret these degrees of freedom
as belonging to a ``dual'' CFT.
The decay spectrum in Eq. (\ref{eq:DecayRate})
should then be proportional to an analytically-continued thermal correlation
function of the CFT, and the QNM frequencies should be related to the
poles of its retarded thermal correlators.

What can we say about the degrees of freedom of this CFT?
A clue comes from Eqs.~(\ref{eq:TiToTH}) and (\ref{eq:WTiWTo}),
and from their formal similarity to Eqs.~(\ref{eq:TLReqTH}) and
(\ref{eq:BTZ_momentum})-(\ref{eq:LRPotentials}).
Consider a pair of thermodynamic systems at temperatures $T_1$ and $T_2$, with chemical
potentials $\mu_1$ and $\mu_2$, coupled to the environment
only through processes where each system changes its internal energy by
the same amount, and similarly for the particle number:  $dU_1 = dU_2$ and
$dN_1 = dN_2$.  Now we view the pair as making up a single combined system,
with $dU = dU_1 + dU_2$ and similarly for $dN$, $dS$,
with $S$ the entropy.
For reversible processes $dS_{1,2} = (1/T_{1,2}) dU_{1,2} + (\mu_{1,2}/T_{1,2}) dN_{1,2}$, so
\begin{equation} \label{revers}
dS = \left(\frac{1}{2 T_1} + \frac{1}{2 T_2}\right) dU + \left( \frac{\mu_1}{2 T_1} + \frac{\mu_2}{2 T_2} \right) dN.
\end{equation}
We interpret this as saying that the combined system
has effectively $T^{-1} = (2T_1)^{-1} + (2 T_2)^{-1}$ and
$\mu/T = \mu_1/2T_1 + \mu_2/2T_2$.  This is just what
we found in Eqs.~(\ref{eq:TiToTH}),(\ref{eq:WTiWTo}), where the two subsystems
are the ones associated with
QNMs and TTMs, and the thermodynamics of the
combined system are just the usual ones expected for
the black hole!
Even the \textit{statistics} of the emitted particles,
determined by the imaginary part of the chemical potential, arise
as a sum of contributions from the two subsystems.
On this basis we propose that the dual description should involve two distinct
sets of degrees of freedom, somehow related to QNMs and TTMs.  Speculations
on partitions of the black hole into two subsystems, involving relations similar to
Eq.~\eqref{revers}, have appeared before in \textit{e.g.} \cite{Wu:2004yk}.

This proposal is similar to what happened in the second and third cases
we reviewed above, where the two subsystems consisted of right- and left-movers
in the CFT, and entered in a symmetrical way
\footnote{The first case [Eq.~(\ref{eq:DecayBPS})] can be understood as $2T_1=T_H$, so
Eq.~(\ref{eq:TLReqTH}) yields $T_2\to \infty$ and one Boltzmann weight is trivial.}.
In our case the two subsystems
are associated with QNMs and TTMs,
and there is no symmetry between them; in particular, the emission spectrum
includes a denominator Boltzmann factor associated with QNMs but none for TTMs.
Perhaps the correct picture here involves a single excitation
associated with QNMs decaying into two quanta, one of which enters the
subsystem associated with TTMs while the other emerges as Hawking radiation.

As argued in \S\ref{sec:Excitations}, QNMs and TTMs are related
to classical bound states along $l_o$ and $l_i$, respectively. This
suggests that the two sets of microscopic degrees of freedom correspond
somehow to $l_o$ and $l_i$, or more generally to geodesics that cross respectively
outside and inside the outer horizon. When $l_o$ and $l_i$ are combined, the
loop formed admits traveling waves which are related to TRMs and therefore
to Hawking radiation.
This pictorially parallels the above
suggestion that microscopic degrees of freedom corresponding to the QNM
and TTM sectors interact to produce Hawking radiation.
It is possible that there is a relation between interactions
among degrees of freedom involved in the production of Hawking radiation
on the microscopic side, and interactions between excitations along $l_o$
and $l_i$ forming loop excitations on the classical side. If so, the classical picture
discussed in \S\ref{sec:Excitations} illustrates
why TTMs are not seen in Eq.~(\ref{eq:DecayRate}), and supports the
notion that the production of a Hawking quantum involves the decay of a
QNM-related quantum into the TTM sector.

The excitations along the contours $l_{i,o}$ are semiclassical, so
heuristically the probability to find an excited quantum at a point $x$,
$P(x) \propto |E_x-V_x(x)|^{-1}$, is inversely proportional to the classical
velocity and substantial only near the turning points $r_{1,2}$.
It is natural to speculate that the relevant dual
description is similarly ``localized'' around those two turning points, by
analogy to the dual descriptions of extremal black holes, which are
localized near the horizon.
Moreover, in our analysis of the resonance
spectrum the starring role was played by complexified geodesics
which connect the two turning points.  This is somewhat reminiscent of the
discussion of asymptotically AdS black holes in \cite{Kraus:2002iv,Fidkowski:2003nf}; there
one has a dual description localized at the two boundaries of the spacetime, and
correlators of very massive scalars between these two boundaries are
dominated by complex geodesics connecting them.  These
correlators in particular determine the massive QNM spectrum.  It would be
interesting to understand whether there is any connection between the two
situations.

\section{Summary and Discussion}
\label{sec:Discussion}

This paper analyzes the spectroscopic properties of a rotating
black hole in the highly-damped frequency regime. More precisely, it is a study of
the evolution of linear perturbations of a massless field with arbitrary
spin, in the spacetime of a four-dimensional rotating, charged (for $s=0$) black hole,
in the large, nearly imaginary frequency range.
Our analysis and main conclusions are as
follows.

\begin{enumerate}

\item
Evidence is presented (in \S\ref{subsec:EquatorialFocusing}) to show that
highly-damped perturbations are equatorially confined, with a
characteristic opening angle $\Delta\theta \sim |m/\omega a|$.
\item
The problem of transmission and reflection is analytically solved
(\S\ref{sec:Analysis}) using the WKB approximation, Stokes phenomenon and
monodromy matching, as illustrated in Figure \ref{fig:Computation}.
The resulting expressions for ${\nT}$ and ${\nR}$ are given in
Eqs.~(\ref{eq:TGeneral})-(\ref{eq:RR}).
\begin{enumerate}
\item
The analysis exploits two complex WKB turning points $r_{1,2}$ and the
steepest-descent (anti-Stokes) lines $l_j$ emanating from them in the
complex $r$-plane, as shown in Figure \ref{fig:Stokes}.
\item
The results depend essentially on two integrals
$S_{o,i}$ [Eqs.~(\ref{eq:SjDefinition}), (\ref{eq:GeodesicAction})] running
along two of these contours, $l_{o,i}$, which cross the real axis respectively
outside the outer event horizon and between the inner and outer horizons.
\item
The points $r_{1,2}$ asymptotically approach complex-conjugate
turning points of small impact parameter null geodesics, in which $\dot{r}=\dot{t}=0$
(\S\ref{subsec:GeneralGeodesics}).
\item
${\nT}$ and ${\nR}$ have poles and zeros corresponding to quasinormal (QNM), total
transmission (TTM) and total reflection (TRM) modes. Their properties are
studied in \S\ref{subsec:BoltzmannFactors} and \S\ref{sec:Excitations}, summarized in Table \ref{tab:modes},
and illustrated in Figure \ref{fig:To}.
\item
${\nT}$ and ${\nR}$ can be written as ratios between three Boltzmann-like weights
$e^{(\omega-\widetilde{\omega}_j)/2 T_j}\pm1$, defined in
Eqs.~(\ref{eq:DefinitionTj})-(\ref{eq:DefinitionWj}) and
related to each other through
Eqs.~(\ref{eq:TiToTH})-(\ref{eq:WTiWTo}).
The frequencies of
each resonant mode are zeros of a corresponding weight.
\end{enumerate}
\item
Each black hole resonance corresponds to a semiclassical bound state
of the Wick-rotated wave equation (\ref{eq:wave_x}) along a
specific contour $l_j$, independent of the boundary conditions at the
horizon and spatial infinity.
\begin{enumerate}
\item
The resonant frequencies [Eqs.~(\ref{eq:resonances_l}),
(\ref{eq:WnUsingGeodesics})] are determined by applying a complexified
Bohr-Sommerfeld equation [(\ref{eq:BS_l}),
(\ref{eq:QuantizedZeroAction})] to the relevant
contour.
\item
The result is $\omega(n) = \widetilde{\omega} + 4\pi i T \left( n+\mu/4
\right)$, where $(4 i T_j)^{-1} = \Delta t_j$ and $\widetilde{\omega}_j \Delta t \propto m$
are respectively the elapsed time and angular position along the corresponding
geodesic [Eqs.~(\ref{eq:TusingGeodesics})-(\ref{eq:zetaDefinition})], and
$\mu$ is a Maslov index.
\item
The QNMs (TTMs) are associated with bound states along $l_o$ ($l_i$),
corresponding to an equatorial ring outside (inside) the outer horizon. The TRMs are
associated with this horizon, and manifest as waves traveling in the
closed loop formed by $l_o$ and $l_i$.
\item
Another contour $l_5$ emanates from a third turning point, encircles the
inner horizon and admits traveling waves similar to the TRMs. These inner
horizon modes (IHMs) are not revealed by ${\nT}$ and ${\nR}$
(\S\ref{subsec:IHMs}).
\end{enumerate}
\item
The results provide hints about the quantum description of the black hole in this
frequency regime.
\begin{enumerate}
\item
The analytically-continued spectrum $\Gamma$ of Hawking radiation escaping the
black hole has the simple form Eq.~(\ref{eq:DecayRate}). It
resembles previously-studied spectra (\S\ref{subsubsec:DualExamples}) which gave
clues to the dual CFT description of black holes.
\item
The relations between QNM and TTM
Boltzmann factors [Eqs.~(\ref{eq:TiToTH})-(\ref{eq:WTiWTo})]
resemble the relations [Eqs.~(\ref{eq:TLReqTH}),
(\ref{eq:BTZ_momentum})-(\ref{eq:LRPotentials})] between the partition functions of ensembles
constructed from two sectors of a dual CFT whose excitations interact to
produce Hawking radiation.
\item
\label{Clause:DualConjecture} We speculate (\S\ref{subsec:DecayRateInterpretation}) that QNMs and TTMs similarly correspond
to distinct sets of microscopic degrees of freedom of some
unknown dual description of the black hole,
which interact to produce Hawking radiation.

\end{enumerate}
\end{enumerate}

Linearized perturbations of a rotating black hole are characterized by two
time scales --- the horizon light-crossing time and the rotation period
--- which are of the same order of magnitude far from the Schwarzschild
and the extremal limits. Analyses of perturbations with a single
time-scale and radiative boundary conditions are complicated by strong
damping. The highly-damped regime studied in this paper is more susceptible
to analytical methods because the decay rate is taken to be much faster
than the characteristic inverse time scale.

In this regime the analysis is simplified by focusing on certain contours
$l_{o,i}$ in the complex $r$-plane. As in previous studies, such contours
play an important role in the WKB analysis. In addition, they provide a
semiclassical, essentially one-dimensional description of the black hole
interactions with its environment.
The black hole resonances can be modeled as
bound states of the Wick-rotated wave equation along $l_{o,i}$, and
scattering off the black hole can be understood in terms of tunneling
between these contours.

In the highly-damped regime, the transmission-reflection amplitudes and
the corresponding resonances are rather insensitive to the details of the potential
barrier surrounding the black hole. Any frequency-independent ``contaminant''
potential may be added to the potential in Teukolsky's equation or to
$V_x, V_z$ without changing any of our results to leading order in
$|\omega|^{-1}$. This frequency regime is universal in the sense that the
results depend simply on (and are formally independent of) $s$, $l$ and
$m$, and are periodic in $\omega_I$.
Moreover, since we keep $\omega_R$ finite, the Boltzmann weights appearing in
the result are finite in this limit, which allows us to determine \textit{e.g.} whether the
denominators correspond to fermionic or bosonic statistics.
In sum, the combination of
analytic results, robustness and universality makes the
highly-damped regime a particularly interesting place.
On the other hand, it is far from clear how one should physically interpret the results
of scattering computations in this regime; here we have only presented
a few speculations in that direction.

The analysis presented here does not directly apply to the Schwarzschild
case $a=0$, where the turning points coalesce to $r=0$, nor to the
extremal case $M^2-a^2-Q^2=0$, where the inner and outer horizons merge to
cutoff $l_2$. It does hold arbitrarily close to these limiting cases
\cite{Keshet:2007nv}. Moreover, much of our discussion extends beyond
the four-dimensional rotating black hole, including for example the
connections between the resonance spectrum and coordinate distances
along geodesics.
In Appendix \ref{subsec:GeneralizedTR} we show how the computations of
$\nT$ and $\nR$ in the highly damped frequency regime can be generalized to
a large class of black hole backgrounds, and demonstrate how the QNM and
TTM conditions may be written in terms of the corresponding geodesics.

Previous studies of the QNMs of the Kerr black hole show
that they fall into two families, only one of which survives to the highly
damped regime \cite{Berti:2004um,Keshet:2007nv}.  It was argued
numerically \cite{Onozawa:1997ux,Berti:2003jh} and analytically \cite{Hod:2005ha}
that the other family of QNM
frequencies approaches $\omega_R = m \Omega$ before disappearing.
Our analysis suggests an explanation for this behavior:
we found that the branch cut in $\nT$ and $\nR$ is naturally placed at
$\omega_R = m \Omega$.  Perhaps the other family of modes
hides behind this cut.

\acknowledgements

We thank J. Maldacena and S. Hod for inspiration and helpful
discussions. U. K. also thanks Y. Harness for helpful advice. This work
was supported by the NSF (grant PHY-0503584). A.~N. is also supported by
the Martin A. and Helen Chooljian Membership at the Institute for Advanced
Study.

\appendix

\section{Generalized transmission-reflection analysis}
\label{subsec:GeneralizedTR}

Here we generalize the computation of highly-damped transmission and reflection
amplitudes $\nT$ and $\nR$ for an arbitrary black hole in which a closed anti-Stokes
contour can be constructed around the (outer) event horizon $r_+$. In particular, this
includes the Schwarzschild and Reissner-Nordstr\"{o}m black holes in
various dimensions. The generalization shows that quite generally
${\nT}(\omega)$ does have zeros, which then must cancel with the poles
of $n_H(\omega)$ as argued in \S\ref{subsec:HawkingCancellation}.

Consider frequencies near the poles of $n_H$, shown in
\S\ref{subsec:HawkingCancellation} to occur when
$e^{4\pi\omega\sigma_+}=1$, where $-\sigma_+$ is the dominant exponent of
$f(r)$ at $r_+$. If $i\omega z(r\simeq r_+) \simeq i\omega\sigma_+\ln(r-r_+)$,
$\omega\sigma_+\in i\R$ is a sufficient condition for the
anti-Stokes lines to avoid $r_+$; it is satisfied near the poles of
$n_H$. Consider the closed contour $C$ obtained by connecting the
anti-Stokes lines closest to $r_+$, denoted $l_0,l_1,l_2,\ldots,l_n=\check{l}_0$
in clockwise (counterclockwise) order for $\omega_R<\omega_c$ ($\omega_R>\omega_c$),
where $\omega_c$ specifies the location of the branch cut.
Along $C$, $f=c_+(l_j) f_+ + {\nT} f_-$, because $f_-$ is dominant inside
$C$ so that $c_-$ can be continued directly to $r_+$. The Stokes
phenomenon implies that $c_+(l_n)=c_+(l_0)\alpha_0^{1/2}-i\varepsilon
{\nT}\alpha_0^{-1/2}\alpha$, where
we defined $\varepsilon\equiv\mbox{sign}(\omega_R-\omega_c)$.
Here, $\alpha\equiv \sum_{k=0}^{n-1} \alpha_k$, where
$\alpha_k=\exp[2i\omega(z_n-z_k)]$ are the relative phases accumulated by
$c_+$ and $c_-$ at the turning points $r_k$, labeled such that $r_k$
follows $l_k$ along $C$. Note that $\alpha_0=\exp(-\varepsilon
4\pi\omega\sigma_+)$. On the other hand the boundary condition at the
horizon implies $c_+(l_n)=\alpha_0^{-1/2}c_+(l_0)$, yielding
\begin{equation} \label{eq:GeneralTUsingLobe}
{\nT} = -i \varepsilon c_+(l_0) \frac{\alpha_0 - 1} {\alpha} \fin
\end{equation}
As $n_H = \pm(\alpha_0-1)^{-1}$, the appearance of zeros of $\nT$ and their cancellation
with the poles of $n_H$ is evident, regardless of the number of turning
points or the associated phases.

The analysis may be pursued further in cases where $f_+$ may be continued
to $r\rightarrow\infty$ such that $c_+(l_1)={\nR}$, as in
the four-dimensional black holes mentioned above. In this case,
Eq.~(\ref{eq:GeneralTUsingLobe}) becomes $n_H {\nT}\alpha=\mp i\epsilon{\nR}$. In
the highly damped regime quite generally $\tT(-\omega)=1$, and at least in
several cases $\tR(-\omega)=i\varepsilon p$ for some constant $p$, so
Eq.~(\ref{eq:ContinuedFluxConservation}) implies that
${\nT}(\omega)+i\varepsilon p {\nR}(\omega)=1$. Combining this with the above
conclusions yields
\begin{equation}
{\nT} = \frac{\alpha_0-1}{\alpha_0-1-p\alpha}
\end{equation}
and
\begin{equation}
{\nR} = i \varepsilon \frac{\alpha}
{\alpha_0-1-p\alpha} \fin
\end{equation}
The QNM and TTM resonant conditions are now identified respectively as
$\alpha_0-1-p\alpha=0$ and
$\alpha=0$.

In the present case of a 4D rotating black hole, $n=2$,
$\alpha_1=e^{2i\varepsilon S_o}$ and $p=1$, reproducing
Eqs.~(\ref{eq:TGeneral}) and (\ref{eq:RGeneral}). For gravitational
perturbations of a 4D
Schwarzschild black hole, for example, one finds \cite{Andersson:2003fh} $n=2$,
$\alpha_1=\alpha_0$, and $p=2$, reproducing the results
of \cite{Neitzke:2003mz}.

In the Schwarzschild and Reissner-Nordstr\"{o}m black holes the
connected system of excitation lines is more complicated than in
the present case of a rotating black hole
and there is no 1-1 correspondence
between resonances and excitation lines.  As in \S\ref{sec:Geodesics}, the
resonances can still be related to the coordinate distance along the associated
geodesics. For example, the condition of \cite{Andersson:2003fh} for
highly-damped gravitational QNMs of a 4D Schwarzschild black hole can be written as
\begin{equation}
1 + 3\exp \left[ \omega \Delta t-\frac{l+1}{2}\Delta\phi + \mbox{spin
term} +\ldots \right] = 0 \coma
\end{equation}
where in this case $\Delta t=\pm1/T_H$, and the
subleading terms in the exponent are all $O(|\omega|^{-1})$.

\bibliography{DampedScattering}

\end{document}